\documentclass[12pt]{article}
\usepackage{achemso,graphics}
\newcommand{\B}[1]{{\mathbf{#1}}}

\title{The AGBNP2 Implicit Solvation Model}

\author{Emilio Gallicchio\thanks{Corresponding author,
emilio@biomaps.rutgers.edu}, Kristina Paris, and Ronald M. Levy\\
Department of Chemistry and Chemical Biology\\
and BioMaPS Institute for Quantitative Biology\\
Rutgers University, Piscataway NJ 08854
}

\begin{document}

\maketitle

\begin{abstract}
The AGBNP2 implicit solvent model, an evolution of the Analytical
Generalized Born plus Non-Polar (AGBNP) model we have previously
reported, is presented with the aim of modeling hydration effects
beyond those described by conventional continuum dielectric
representations. A new empirical hydration free energy component based
on a procedure to locate and score hydration sites on the solute
surface is introduced to model first solvation shell effects, such as
hydrogen bonding, which are poorly described by continuum dielectric
models.  This new component is added to the Generalized Born and
non-polar AGBNP models which have been improved with respect to the
description of the solute volume description. We have introduced an
analytical Solvent Excluding Volume (SEV) model which reduces the
effect of spurious high-dielectric interstitial spaces present in
conventional van der Waals representations of the solute volume. The
AGBNP2 model is parametrized and tested with respect to experimental
hydration free energies of small molecules and the results of explicit
solvent simulations. Modeling the granularity of water is one of the
main design principles employed for the the first shell solvation
function and the SEV model, by requiring that water locations have a
minimum available volume based on the size of a water molecule. We
show that the new volumetric model produces Born radii and surface
areas in good agreement with accurate numerical evaluations. The
results of Molecular Dynamics simulations of a series of mini-proteins
show that the new model produces conformational ensembles in
substantially better agreement with reference explicit solvent
ensembles than the original AGBNP model with respect to both
structural and energetics measures.
\end{abstract}

\section{Introduction}

Water plays a fundamental role in virtually all biological processes.
The accurate modeling of hydration thermodynamics is therefore
essential for studying protein conformational
equilibria,
%\cite{Prabhu:Sharp:2006} 
aggregation,
%\cite{Shea:Berne} 
and binding.
%\cite{} 
Explicit
solvent models arguably provide the most detailed and complete
description of hydration phenomena.\cite{Levy:Gallicchio:98} They are,
however, computationally demanding not only because of the large
number of solvent atoms involved, but also because of the need to
average over many solvent configurations to obtain meaningful
thermodynamic data.  Implicit solvent models,\cite{Feig:Brooks:Review:2004}
which are based on the statistical mechanics concept of the solvent
potential of mean force,\cite{Roux:Simonson:99} have been shown to be
useful alternatives to explicit solvation for applications including
protein folding and binding,\cite{Felts:Andrec:Gallicchio:Levy:2008}
and small molecule hydration free energy
prediction.\cite{Gallicchio:Zhang:Levy:2002}

Modern implicit solvent
models\cite{Onufriev:2008,Chen:Brooks:Khandogin:2008} include distinct
estimators for the non-polar and electrostatic components of the
hydration free energy. The non-polar component corresponds to the free
energy of hydration of the uncharged solute while the electrostatic
component corresponds to the free energy of turning on the solute
partial charges. The latter is typically modeled
treating the water solvent as a uniform high-dielectric
continuum.\cite{Tomasi:Persico:94} Methods based on the numerical
solution of the Poisson-Boltzmann (PB) equation\cite{Baker:2005}
provide a virtually exact representation of the response of the
solvent within the dielectric continuum approximation.  Recent
advances extending dielectric continuum approaches have focused on the
development of Generalized Born (GB) models,\cite{Still:90} which have
been shown to reproduce with good accuracy PB and explicit
solvent\cite{Zhang:Gallicchio:Friesner:Levy:2001,Chen:Brooks:Khandogin:2008}
results at a fraction of the computational expense.  The development
of computationally efficient analytical and differentiable GB methods based on pairwise descreening
schemes\cite{Schaefer:Froemmel:90,Hawkins:Cramer:Truhlar:96,Onufriev:2008}
has made possible the integration of GB models in molecular dynamics
packages for biological
simulations.\cite{Dominy:Brooks:99,Banks:Gallicchio:Levy:2005,Case:AmberReview:2005}

The non-polar hydration free energy component accounts for all
non-electrostatic solute-solvent interactions as well as hydrophobic
interactions,\cite{Ben-Naim:80} which are essential driving forces in
biological processes such as protein
folding\cite{Kauzmann:59,Dill:90,Privalov:Makhatadze:93,Honig:Yang:95}
and
binding.\cite{Sturtevant:77,Williams:Searle:Mackay:Gerhard:Maplestone:93,Froloff:Windemuth:Honig:97,Siebert:Hummer:2002}
Historically the non-polar hydration free energy has been modeled by
empirical surface area models\cite{Ooi:Oobatake:Nemethy:Sheraga:87}
which are still widely employed.\cite{Lee:Duan:Kollman:2000,Hunenberger:Helms:Narayana:Taylor:McCammon:99,Simonson:Brunger:94,Sitkoff:Sharp:Honig:94b,Still:90,Rapp:Friesner:99,Fogolari:Esposito:Viglino:Molinari:2001,Pellegrini:Field:2002,Curutchet:Cramer:Truhlar:2003,Jorgensen:Ulmschneider:Tirado-Rives:2004}
Surface area models are useful as a first approximation, however
qualitative deficiencies have been observed.\cite{Simonson:Brunger:94,Wallqvist:Covell:95,Gallicchio:Kubo:Levy:2000,Levy:Zhang:Gallicchio:Felts:2003,Wagoner:Baker:2006,Chen:Brooks:2008,Mobley:Bayly:Dill:2009} 

Few years ago we presented the Analytical Generalized Born plus Non-Polar
(AGBNP) implicit solvent model,\cite{Gallicchio:Levy:2004}
which introduced two
key innovations with respect to both the electrostatic and non-polar
components. Unlike most implicit solvent models, the AGBNP non-polar
hydration free energy model 
includes distinct estimators for the solute--solvent van der Waals
dispersion energy and cavity formation work components. The main
advantages of a model based on the cavity/dispersion decomposition of
the non-polar solvation free energy stem from its ability to describe
both medium range solute--solvent dispersion interactions, which
depend on solute composition, as well as conformational equilibria
dominated by short-range hydrophobic interactions, which mainly depend
on solute conformation alone.\cite{Chen:Brooks:2008} A series of
studies highlight the importance of the balance between hydrophobicity and
dispersion interactions in regulating the structure of the hydration
shell and the strength of interactions between macromolecules.\cite{Wallqvist:Gallicchio:Levy:2001,Huang:Chandler:2002,Zhou:Huang:Margulis:Berne:2004} 
In AGBNP the work of
cavity formation is described by a surface area-dependent
model,\cite{Pierotti:76,Hummer:Garde:Garcia:Paulaitis:Pratt:98,Lum:Chandler:Weeks:99,Gallicchio:Kubo:Levy:2000}
while the dispersion estimator is based on the integral of van der
Waals solute--solvent interactions over the solvent modeled as a
uniform continuum.\cite{Levy:Zhang:Gallicchio:Felts:2003} This form of
the non-polar estimator had been motivated by a series of earlier
studies\cite{Pitarch:Moliner:Ahuir:Silla:Tunon:96,Ashbaugh:Kaler:Paulaitis:99,Gallicchio:Kubo:Levy:2000,Gallicchio:Zhang:Levy:2002,Pitera:vanGunsteren:2001,Zacharias:2003}
and has since been shown by
us\cite{Levy:Zhang:Gallicchio:Felts:2003,Su:Gallicchio:Levy:2004,Felts:Harano:Gallicchio:Levy:2004,Felts:Gallicchio:Levy:2008}
and
others\cite{Wagoner:Baker:2006,Dong:Wagoner:Baker:2008,Chen:Brooks:2008,Mobley:Bayly:Dill:2009}
to be qualitatively superior to models based only on the surface area
in reproducing explicit solvent results as well as rationalizing structural and
thermodynamical experimental observations.

The electrostatic solvation model in AGBNP is based on the pairwise
descreening GB scheme\cite{Hawkins:Cramer:Truhlar:96} whereby the Born
radius of each atom is obtained by summing an appropriate descreening
function over its neighbors.  The main distinction between the AGBNP
GB model and conventional pairwise descreening implementations is that
in AGBNP the volume scaling factors, which offset the overcounting of
regions of space occupied by more than one atom, are computed from the
geometry of the molecule rather than being introduced as
geometry-independent parameters fit to either experiments or to
numerical Poisson-Boltzmann
results.\cite{Schaefer:Karplus:96,Qiu:Shenkin:Hollinger:Still:97,Dominy:Brooks:99,Tsui:Case:2000a}
The reduction of number of parameters achieved with this strategy
improves the transferability of the model to unusual functional groups
often found in ligand molecules, which would otherwise require the
introduction of numerous
parameters.\cite{Schaefer:Bartels:Leclerc:Karplus:2001}

Given its characteristics, the AGBNP model has been mainly targeted to
applications involving molecular dynamics canonical conformational
sampling, and to the study of protein-ligand complexes.
Since its inception the model has been employed in the investigation
of a wide variety of biomolecular problems ranging from peptide
conformational propensity prediction and
folding,\cite{Felts:Harano:Gallicchio:Levy:2004,Chekmarev:Ishida:Levy:2004,Andrec:Felts:Gallicchio:Levy:2005,Gallicchio:Andrec:Felts:Levy:2005}
ensemble-based interpretation of NMR
experiments,\cite{Weinstock:Narayanan:Baum:Levy:2007,Weinstock:Narayanan:Baum:Levy:2008}
protein loop homology modeling,\cite{Felts:Gallicchio:Levy:2008} the
study of intrinsically disordered proteins,\cite{Wu:Narayan:Weinstock:Levy:Baum:2009}
ligand-induced conformational changes in
proteins,\cite{Ravindranathan:Gallicchio:Levy:2005,Messina:Talaga:2007}
conformational equilibria of protein-ligand
complexes,\cite{Ravindranathan:Gallicchio:Levy:2006,Ravindranathan:Gallicchio:2007}
protein-ligand binding affinity
prediction,\cite{Su:Gallicchio:Levy:2007} and structure-based vaccine
design.\cite{Lapelosa:Gallicchio:Arnold:Levy:2009} The AGBNP model has
been re-implemented and adopted with minor modifications by other
investigators.\cite{Tjong:Zhou:2007a,Tjong:Zhou:2007b} The main
elements of the AGBNP non-polar and electrostatic models have been
independently
validated,\cite{Zhu:Alexov:Honig:2005,Fan:Mark:Zhu:Honig:2005,Wagoner:Baker:2006,Chen:Brooks:2008}
and have been incorporated in recently proposed hydration free energy
models.\cite{Grant:Pickup:Sykes:Kitchen:Nicholls:2007,Labute:2008}

In this work we present a new implicit solvent model named AGBNP2
which builds upon the original AGBNP implementation (hereafter
referred to as AGBNP1) and improves it with respect to the the
description of the solute volume and the treatment of short--range
solute--water electrostatic interactions. 

Continuum dielectric models assume that the solvent can be described
by a linear and uniform dielectric
medium.\cite{Levy:Belhadj:Kitchen:1991} This assumption is generally
valid at the macroscopic level, however at the molecular level water
exhibits significant deviations from this
behavior.\cite{Levy:Gallicchio:98} Non-linear dielectric responses,
the non-uniform distribution of water molecules, charge asymmetry and
electrostriction effects\cite{Alper:Levy:90} are all phenomena
originating from the finite size and internal structure of water
molecules as well as their specific
interactions which are not taken into account by continuum dielectric models. Some of these effects are qualitatively captured by
standard classical fixed-charge explicit water models, however others,
such as polarization and hydrogen bonding interactions, can be fully
modeled only by adopting more complex physical
models.\cite{Morozov:Kortemme:2005} GB models make further
simplifications in addition to the dielectric continuum approximation,
thereby compounding the challenge of achieving with GB-based implicit
solvent models the level of realism required to reliably model
phenomena, such as protein folding and binding, characterized by
relatively small free energy changes. 

In the face of these challenges a
reasonable approach is to adopt an empirical hydration free energy
model motivated by physical arguments\cite{Lazaridis:Karplus:2000}
parametrized with respect to experimental data on small molecule
hydration.\cite{Privalov:Makhatadze:93} 
Models of this kind typically score
conformations on the basis of the degree of solvent exposure of solute
atoms. Historically\cite{Eisenberg:McLachlan:86} the solvent accessible
surface area of the solute has been used for this purpose, while
modern implementations suitable for conformational sampling applications
often employ computationally convenient volumetric
measures.\cite{Lazaridis:Karplus:99p,Vitalis:Pappu:2008}
In this work we take this general
approach but we retain the GB model component which we assume to form
a useful baseline
to describe the long--range influence of the water medium. The
empirical parametrized component of the model then takes the form of an empirical first solvation shell correction function
designed so as to absorb hydration effects not accurately described by the
GB model. Specifically, as described below, we employ
a short--range analytical hydrogen bonding
correction function based on the degree of water occupancy taking into
account the finite size of water molecules,   of
appropriately chosen hydration sites on the solute
surface. The aim of
this model is to effectively introduce some explicit solvation
features without actually adding water molecules to the system as for example
done in hybrid explicit/implicit models.\cite{Yu:Jacobson:Friesner:2004,Okur:Wickstrom:Simmerling:2008}  

In this work we also enhance the volume description of the solute,
which in AGBNP1 is modeled by means of atomic spheres of radius equal
to the atomic van der Waals radius. The deficiencies of the van der
Waals solute volume model have been
recognized.\cite{Lee:Salsbury:Brooks:2002} They stem from the presence
of high-dielectric interstitial spaces in the solute interior which
are too small to contain discrete water molecules. These spurious high
dielectric spaces contribute to the hydration of buried or partially
buried atoms causing underestimation of desolvation effects.  The
volume enclosed by the molecular surface (MS), defined as the surface
produced by a solvent spherical probe (of $1.4$ \AA\ radius for water)
rolling on the van der Waals surface of the
solute,\cite{Lee:Richards:1971} represents the region which is
inaccessible to water molecules and is often referred as the Solvent
Excluding Volume (SEV).\cite{Pascual:Silla:90} The SEV, lacking the spurious high-dielectric interstitial spaces, provides a
better representation of the low-dielectric region associated with the
solute.  For this reason accurate Poisson-Boltzmann
solvers,\cite{Cortis:Friesner:97,Rocchia:Sridharan:Nicholls:Alexov:Chiabrera:Honig:2002,Baker:2005}
have employed the SEV description of the solute region.

Despite its clear advantages, the lack of of analytical and
computational efficient descriptions of the SEV have hampered its
deployment in conjunction with GB models for molecular dynamics
applications. The GBMV series of
models\cite{Lee:Salsbury:Brooks:2002,Lee:Feig:Salsbury:Brooks:2003,Chocholousova:Feig:2006}
achieve high accuracy relative to numerical Poisson calculations in
part by employing the SEV description of the solute volume. The
analytical versions of
GBMV\cite{Lee:Feig:Salsbury:Brooks:2003,Chocholousova:Feig:2006}
describe the SEV volume by means of a continuous and differentiable
solute density function which is integrated on a grid to yield atomic
Born radii. In this work we present a model for the SEV that preserves
the analytical pairwise atomic descreening approach employed in the
AGBNP1 model,\cite{Gallicchio:Levy:2004} which avoids computations on
a grid.  We show that this approximate model reproduces all of the key
features of the SEV while yielding the same favorable algorithmic scaling of
pairwise descreening approaches.

\section{Methods}

\subsection{The Analytical Generalized Born plus Non-Polar Implicit
  Solvent Model (AGBNP)}

In this section we briefly review the formulation of the AGBNP1
implicit solvent model, which forms the basis for the new AGBNP2
model. A full account can be found in the original
reference.\cite{Gallicchio:Levy:2004} The AGBNP1 hydration free energy
$\Delta G_{\rm h}$ is defined as
\begin{eqnarray}
\Delta G_{\rm h} & = & \Delta G_{\rm elec} + \Delta G_{\rm np}
          \nonumber \\
          &=& \Delta G_{\rm elec} + \Delta G_{\rm cav} + \Delta G_{\rm
          vdW},
\label{hG1}
\end{eqnarray}
where $\Delta G_{\rm elec}$ is the electrostatic contribution to the
solvation free energy and $\Delta G_{\rm np}$ includes
non-electrostatic contributions. $\Delta G_{\rm np}$ is further
decomposed into a cavity hydration free energy $\Delta G_{\rm cav}$
and a solute--solvent van der Waals dispersion interaction component
$\Delta G_{\rm vdW}$. 

\subsubsection{Geometrical estimators}

Each free energy component in Eq.\ (\ref{hG1}) is ultimately based on a analytical
geometrical description of the solute according to which the solute volume
is modeled as a set of overlapping atomic spheres of radii $R_i$
centered on the atomic positions $\B{r}_i$. Hydrogen atoms do not
contribute to the solute volume. The solute volume is computed using
the Poincar\'{e} formula (also known as the inclusion-exclusion
formula) for the volume of the union of a set of intersecting
elements
\begin{equation}
V = \sum_i V_i - \sum_{i<j} V_{ij} + \sum_{i<j<k} V_{ijk} - \ldots  \, 
\label{V1}
\end{equation}
where $V_i = 4\pi R_i^3/3$ is the volume of atom $i$, $V_{ij}$ is the
volume of intersection of atoms $i$ and $j$ (second order
intersection), $V_{ijk}$ is the volume of intersection of atoms $i$,
$j$, and $k$ (third order intersection), and so on. The overlap
volumes are approximated by the overlap integral, $V^{\rm g}_{12 \ldots n}$,
available in analytical form, between $n$ Gaussian density functions each
corresponding to a solute atom:
\begin{equation}
V^{\rm g}_{12 \ldots n} \simeq \int d^3\B{r} \, \rho_1(\B{r}) \rho_2(\B{r}) \ldots \rho_n(\B{r})
\, ,
\label{gauss_ovol}
\end{equation}
where the Gaussian density function for atom $i$ is
\begin{equation}
\rho_i(\B{r}) = p \exp\left[ - c_i  (\B{r}-\B{r}_i)^2 \right]
\, ,
\label{gauss_voli}
\end{equation}
where 
\begin{equation}
c_i = \frac{\kappa}{R_i^2} \, ,
\label{galpha_def}
\end{equation}
and
\begin{equation}
p  = \frac{4 \pi}{3} \left( \frac{\kappa}{\pi} \right)^{3/2} \, ,
\label{pkappa_relation}
\end{equation}
and $\kappa$ is a dimensionless parameter that regulates the
diffuseness of the atomic Gaussian function. In the AGBNP1 formulation $\kappa=2.227$. 

Since Gaussian integrals are in principle non-zero
for any finite distances between the Gaussian densities, the question
arises of how to implement in practice Eq.\ (\ref{V1}) without having to
compute the combinatorial large number of mostly small overlap
volumes between all of the atoms of the molecule. Although
not mentioned in reference \cite{Gallicchio:Levy:2004}, this problem
had been addressed in AGBNP1 by introducing a switching
function that smoothly reduces to zero the overlap volume between two or
more Gaussians when the overlap volume is smaller than a certain
value. If $V^{\rm g}_{12\ldots}$ is the value of the Gaussian overlap
volume between a set of atoms, the corresponding overlap volume
$V_{12\ldots}$ used in Eq.\ (\ref{V1}) is set as
\begin{equation}
V_{12\ldots n} = \left\{ \begin{array}{ll}
0 & V^{\rm g}_{12\ldots} \le v_1 \\
V^{\rm g}_{12\ldots n} f_w(u) & v_1 < V^{\rm g}_{12\ldots n} < v_2  \\
V^{\rm g}_{12\ldots n} & V^{\rm g}_{12\ldots} \ge v_2
\end{array} \right. \, ,
\label{switchA}
\end{equation}
where
\begin{equation}
u = \frac{V^{\rm g}_{12\ldots} - v_1}{v_2 - v_1} \, ,
\end{equation}
\begin{equation}
f_w(x) = x^3 ( 10 - 15 x + 6 x^2) \, ,
\label{switch2}
\end{equation}
and $v_1 = 0.2$ \AA$^3$ and $v_2 = 2$ \AA$^3$. This effectively sets
to zero Gaussian overlap volumes smaller than $v_1$, leaves volumes
above $v_2$ unchanged and smoothly reduces volumes in between
these two limits. This scheme drastically reduces the number of
overlap volumes that need to be calculated since the fact that an
$n$-body overlap volume $V_{12\ldots n}$ between $n$ atoms is zero
guarantees that all of the $n+1$-body overlap volumes corresponding to
the same set of atoms plus one additional atom are also zero.
(Note below that the formulation of
AGBNP2 employs smaller values of $v_1$ and $v_2$ to improve the
accuracy of surface areas).

The van der Waals surface area $A_i$ of atom $i$, which is another geometrical
descriptor of the model, is
based on the derivative $\partial V/\partial R_i$
of the solute volume with respect to the radius $R_i$\cite{Kratky:81}
\begin{equation}
A_i = f_a\left( \frac{\partial V}{\partial R_i} \right)
\label{Aidef}
\end{equation}
where $V$ is given by Eq.\ (\ref{V1}) and
\begin{equation}
f_a(x) = \left\{
\begin{array}{lc}
\frac{x^3}{a^2 + x^2} & x > 0 \\
0 & x \le 0
\end{array}
\right. \, .
\label{ai_filter}
\end{equation}
with $a = 5$ \AA$^2$, is a filter function which prevents negative
values for the surface areas for buried atoms while inducing
negligible changes to the surface areas of solvent-exposed atoms.

The model further defines the so-called self-volume $V'_i$ of atom $i$ as
\begin{equation}
V'_i = V_i - \frac12 \sum_j V_{ij} + \frac13 \sum_{j<k} V_{ijk} +
\ldots \, .
\label{Vself}
\end{equation}
which is computed similarly to the solute volume and measures the
solute volume that is considered to belong exclusively to
this atom. Due to the overlaps with other atoms, the self-volume
$V'_i$ of an atom is smaller than the van der Waals volume $V_i$ of the atom.
The ratio
\begin{equation}
s_i = \frac{V'_i}{V_i} \le 1
\label{sidef}
\end{equation}
is a volume scaling factor is used below in the evaluation of the Born
 radii.

\subsubsection{Electrostatic Model}

The electrostatic hydration free energy is modeled using a continuous
dielectric representation of the water solvent using
the Generalized Born (GB) approximation
\begin{equation}
\Delta G_{\rm elec} = u_\epsilon \sum_{i} \frac{q_i^2}{B_i} + 
2 u_\epsilon \sum_{i<j} \frac{q_i q_j}{f_{ij}}  \, ,
\label{Ggbdef_sep}
\end{equation}
where
\begin{equation}
u_\epsilon = -\frac12 \left( \frac{1}{\epsilon_{\rm in}} -
\frac{1}{\epsilon_{\rm w}}\right) \, ,
\label{uedef}
\end{equation}
and where $\epsilon_{\rm in}$ is the dielectric constant of the interior
of the solute, $\epsilon_{\rm w}$ is the dielectric constant of the
solvent; $q_i$ and $q_j$ are the charges of atom $i$ and $j$, and
\begin{equation}
f_{ij} = \sqrt{r_{ij}^2 + B_i B_j \exp(-r_{ij}^2/4 B_i B_j)} \, .
\label{fgbdef}
\end{equation}
In Eqs.\ (\ref{Ggbdef_sep})--(\ref{fgbdef}) $B_i$ denotes the
Born radius of atom $i$ which, under the Coulomb field
approximation, is given by the inverse of the integral over
the solvent region of the
negative 4th power of the distance function centered on
atom $i$,
\begin{equation}
\beta_i = 1/B_i = \frac{1}{4 \pi}
\int_{\rm Solvent} d^3\B{r} \, \frac{1}{( \B{r} - \B{r}_i)^4} \, .
\label{Bintegral}
\end{equation}
In the AGBNP1 model this integral is approximated by a so-called 
pairwise descreening formula
\begin{equation}
\beta_i  = \frac{1}{R_i} - \frac{1}{4 \pi} \sum_{j \ne i} s_{ji} Q_{ji}
\, ,
\label{Bpairwise_sji} 
\end{equation}
where $R_i$ is the van der Waals radius of atom $i$, $s_{ji}$ is the
volume scaling factor for atom $j$ [Eq.\ (\ref{sidef})] when atom
$i$ is removed from the solute and $Q_{ji}$ is the integral
(available in analytic form see Appendix X of reference X) of the
function $( \B{r} - \B{r}_i)^{-4}$ over the volume of the sphere
corresponding to solute atom $j$ that lies outside the sphere
corresponding to atom $i$.  Eq.\ (\ref{Bpairwise_sji}) applies to all
of the atoms $i$ of the solute (hydrogen atoms and heavy atoms)
whereas the sum over $j$ includes only heavy atoms.  The AGBNP1
estimate for the Born radii $B_i$ are finally computed from the
inverse Born radii $\beta_i$ from Eq.~(\ref{Bpairwise_sji}) after
processing them through the function:
\begin{equation}
B_i^{-1} = f_b(\beta_i) = \left\{
\begin{array}{lc}
\sqrt{b^2 + \beta_i^2} & \beta_i > 0 \\
b & \beta_i \le 0
\end{array}
\right. \, ,
\label{bi_filter}
\end{equation}
where $b^{-1} = 50$ \AA. The filter function Eq~(\ref{bi_filter}) is
designed to prevent the occurrence of negative Born radii or Born
radii larger than $50$ \AA. The goal of the filter function is simply
to increase the robustness of the algorithm in limiting cases.  The
filter function has negligible effect for the most commonly observed
Born radii smaller than $20$ \AA.

In the AGBNP1 model the scaling factors $s_{ji}$ are
  approximated by the expression
\begin{equation}
s_{ji} \simeq s_j + \frac12 \frac{V_{ij}}{V_j} \, ,
\label{spijdef}
\end{equation}
where $s_j$ is given by Eq.\ (\ref{sidef}) and $V_{ij}$ is the two
body overlap volume between atoms $i$ and $j$. Also, in the original
AGBNP formulation the computation of the scaling factors and the
descreening function in Eq.\ (\ref{Bpairwise_sji}) employed the van
der Waals radii for the atoms of the solute and the associated
Gaussian densities. These two aspects have been modified in the new
formulation (AGBNP2) as described below.

\subsubsection{Non-Polar Model}

The non-polar hydration free energy is decomposed into the cavity
hydration free energy $\Delta G_{\rm cav}$ and the solute--solvent van
der Waals dispersion interaction component $\Delta G_{\rm vdW}$.  The
cavity component is described by a surface area
model\cite{Pierotti:76,Hummer:Garde:Garcia:Paulaitis:Pratt:98,Lum:Chandler:Weeks:99,Gallicchio:Kubo:Levy:2000}
\begin{equation}
\Delta G_{\rm cav} = \sum_i \gamma_i A_i  \, ,
\label{Gcav}
\end{equation}
where the summation runs over the solute heavy atoms, $A_i$ is the van
der Waals surface area of atom $i$ from Eq.\ (\ref{Aidef}), and
$\gamma_i$ is the surface tension parameter assigned to atom $i$ (see
Table X of reference X). Surface areas are computed using augmented
radii $R^{\rm c}_i$ for the atoms of the solute and the associated Gaussian
densities. Augmented radii are defined as the the van der Waals radii
(Table X of reference X) plus a $0.5$ \AA\ offset.
The computation of the atomic surface areas in AGBNP2 is mostly unchanged from
the original implementation,\cite{Gallicchio:Levy:2004} with the exception of the
values of the switching function cutoff parameters $v_1$ and $v_2$ of Eq.\
(\ref{switchA}), which in the new model are set as  $v_1 = 0.01$
\AA$^3$ and $v_2 = 0.1$ \AA$^3$. This change was deemed necessary to
improve the accuracy of the surface areas which in the new model also
affect the Born radii estimates through Eq.\ (\ref{sj_def}) below.

The solute-solvent van der Waals free energy term is  modeled by the expression
\begin{equation}
\Delta G_{\rm vdW} = \sum_i \alpha_i \frac{a_i}{(B_i + R_w)^3} \, ,
\label{Gvdw_def}
\end{equation}
where $\alpha_i$ is an adjustable dimensionless parameter on the order
of 1 (see Table X of reference X) and
\begin{equation}
a_i = - \frac{16}{3} \pi \rho_w \epsilon_{iw} \sigma_{iw}^6 \, ,
\end{equation}
where $\rho_w = 0.033428$ \AA$^{-3}$ is the number density of water at
standard conditions, and $\sigma_{iw}$ and $\epsilon_{iw}$ are the
OPLS force field\cite{Jorgensen:Maxwell:96} Lennard-Jones interaction parameters for the
interaction of solute atom $i$ with the oxygen atom of the TIP4P water
model.\cite{Jorgensen:Madura:85} If $\sigma_i$ and $\epsilon_i$ are the OPLS
Lennard-Jones parameters for atom $i$,
\begin{eqnarray}
\sigma_{iw}   & = & \sqrt{\sigma_i \sigma_w} \\
\epsilon_{iw} & = & \sqrt{\epsilon_i \epsilon_w} \, ,
\end{eqnarray}
where $\sigma_w = 3.15365$ \AA, and $\epsilon_w = 0.155$ kcal/mol are
the Lennard-Jones parameters of the TIP4P water oxygen. In
Eq.~(\ref{Gvdw_def}) $B_i$ is the Born radius of atom $i$ from Eqs.\
(\ref{Bpairwise_sji}) and (\ref{bi_filter}) and $R_w =
1.4$ \AA\ is
a parameter corresponding to the radius of a water
molecule.

\subsection{Pairwise Descreening Model using the Solvent Excluding Volume}

\begin{figure}[tbp]
\begin{center}
\scalebox{0.75}{\includegraphics{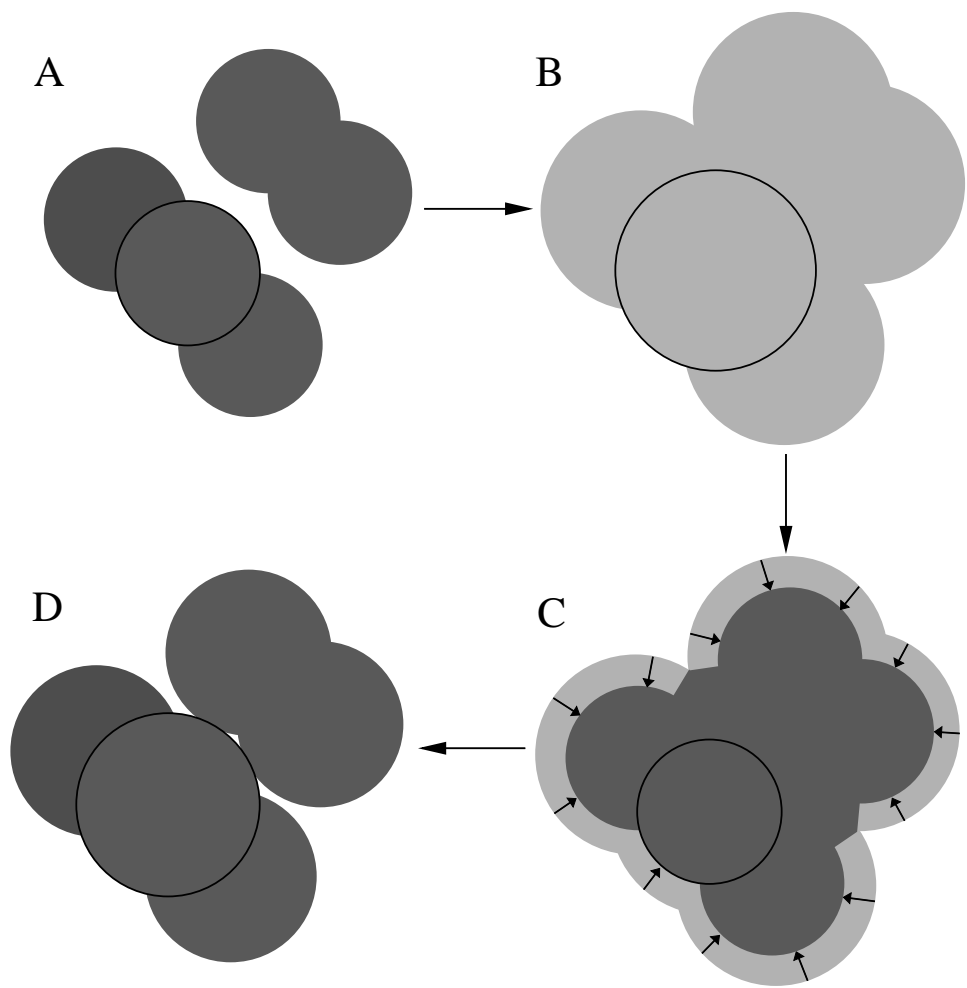}}
\end{center}
\caption{\label{spheres-ms.fig} 
{\small
Schematic diagram illustrating the
ideas underpinning the model for the solvent excluded volume
descreening. Circles represent atoms of two idealized solutes placed
in proximity of each other. The van der Waals description of the
molecular volume (panel A) leaves high dielectric interstitial spaces
that are too small to fit water molecules. The adoption of enlarged
van der Waals radii (B) removes the interstitial spaces but
incorrectly excludes solvent from the surface of solvent exposed
atoms. The solvent volume subtended by the solvent-exposed surface
area is subtracted from the enlarged volume of each atom (C) such that
larger atomic descreening volumes are assigned to buried atoms
(circled) than exposed atoms (D), leading to the reduction of
interstitial spaces while not overly excluding solvent from the
surface of solvent exposed atoms.  
}
}
\end{figure}

\begin{figure}[tbp]
\begin{center}
\scalebox{0.75}{\includegraphics{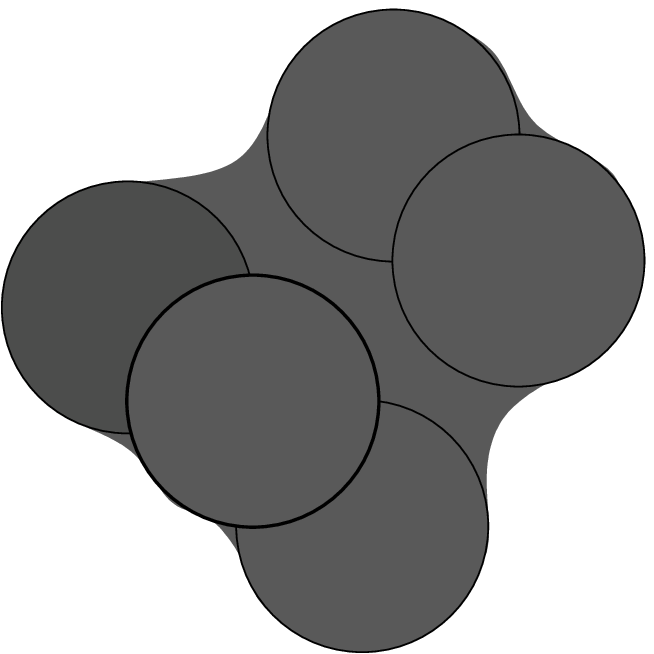}}
\end{center}
\caption{\label{ms.fig} 
{\small
Illustration of the
relationship between the van der Waals volume and the solvent excluded
volume enclosed by the molecular surface.  
}
}
\end{figure}

When using van der Waals radii to describe the solute volume, small
crevices between atoms (Fig. \ref{spheres-ms.fig}, panel A) are
incorrectly considered as high-dielectric solvent
regions,\cite{Onufriev:Bashford:Case:2000,Lee:Feig:Salsbury:Brooks:2003,Mongan:Simmerling:McCammon:Case:Onufriev:2007}
leading to underestimation of the Born radii, particularly for buried
atoms. The van der Waals volume description implicitly ignores the
fact that the finite size of water molecules prevents them to occupy
sites that, even though are not within solute atoms, are too small to
be occupied by water molecules. Ideally a model for the Born radii
would include in the descreening calculation all of the volume
excluded from water either because it is occupied by a solute atom or
because it is located in an interstitial region inaccessible to water
molecules. We denote this volume as the Solvent Excluding Volume
(SEV).  A realistic description of the SEV is the volume enclosed
within the molecular surface\cite{Lee:Richards:1971} of the solute obtained by
tracing the surface of contact of a sphere with a radius characteristic
of a water molecule (typically 1.4 \AA) rolling over the van der Waals
surface of the solute. The main characteristic of this definition of
the SEV (see Fig \ref{ms.fig}) is that, unlike the van der Waals
volume, it lacks small interstitial spaces while it closely resembles
the van der Waals volume near the solute-solvent interface.  The
molecular surface description of the SEV can not be easily implemented
into an analytical formulation. In this section we will present an
analytical description of the SEV for the purpose of the pairwise
descreening computation of the Born radii as implemented in AGBNP2
that preserves the main characteristics of the molecular surface
description of the SEV.

The main ideas underpinning the SEV model presented here are
illustrated in Fig.\ \ref{spheres-ms.fig}. We start with the van der
Waals representation of the solute (model A) which presents an
undesirable high dielectric interstitial space between the two groups
of atoms. Increasing the atomic radii leads to a representation (model
B) in which the interstitial space is removed but that also
incorrectly excludes solvent from the surface of solvent exposed
atoms. This representation is therefore replaced with one in which the
effective volume of each atom in B is reduced by the volume subtended
between the solvent-exposed surface of each atom and its van der Waals
radius (panel C). This process yields model D in which the effective volume of
the most buried atom is larger than those of the solvent-exposed
atoms. This SEV model covers the interstitial high-dielectric spaces
present in a van der Waals description of the solute volume, while
approximately maintaining the correct van der Waals volume description
of atoms at the solute surface as in the molecular surface description
of the SEV (Fig.\ \ref{ms.fig}).

These ideas have been implemented in the AGBNP2 model as follows. The
main modification consists of adopting for the pairwise descreening
Generalized Born formulation the same augmented van der Waals radii as
in the computation of the atomic surface areas. As in the previous
model the augmented atomic radii, $R^c_i$, are set as
\begin{equation}
R^{c}_i = R_i + \Delta R \, ,
\label{big_radii}
\end{equation}
where $R_i$ is the van der Waals radius of the atom and $\Delta R =
0.5$ \AA\ is the offset. The augmented radii are used in the same way
as in the AGBNP1 formulation to define the atomic spheres and the
associated Gaussian densities
[Eqs.\ (\ref{gauss_ovol})--(\ref{pkappa_relation})]. Henceforth in
this work all of the quantities (atomic volumes, self volumes, etc.)
are understood to be computed with the augmented atomic radii, unless
otherwise specified.  In AGBNP2 the form of the expression for the
inverse Born radii [Eq.\ (\ref{Bpairwise_sji})] is unchanged however
the expressions for the volume scaling factors $s_{ji}$ and the
evaluation of the descreening function $Q_{ji}$ are modified as
follows to introduce the augmented atomic radii and the reduction of
the atomic self volumes in proportion to the solvent accessible
surface areas as discussed above.

\begin{figure}[tbp]
\begin{center}
\includegraphics{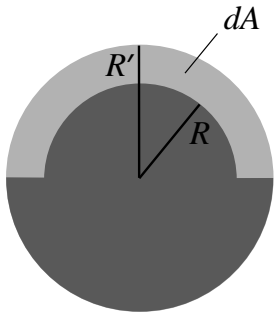}
\end{center}
\caption{\label{surfvol.fig} {\small
Graphical construction showing the volume subtracted from the
atomic self-volume to obtain the surface-area corrected atomic
self-volume. $R$ is the van der Waals radius of the atom, $R' = R +
\Delta R$ is the enlarged atomic radius. $dA$ is the volume of the
region (light gray) subtended by the solvent-exposed surface area in
between the enlarged and van der Waals atomic spheres.  
}
}
\end{figure}

The pairwise volume scaling factors $s_{ji}$, that is the volume
scaling factor for atom $j$ when atom $i$ is removed from the solute,
are set as 
\begin{equation}
s_{ji} = s_j + \frac{V'_{ji}}{V_j}
\label{new_sji}
\end{equation}
where
\begin{equation}
s_j = \frac{V'_j - d_j A_j}{V_j} \, 
\label{sj_def}
\end{equation}
is the intrinsic scaling factor for atom $j$ (computed with all the
atoms present), and the quantity
\begin{equation}
V'_{ji} = V'_{ij} = \frac12 V_{ij} - \frac13 \sum_k V_{ijk} +
\frac14 \sum_{k<l} V_{ijkl} - \ldots \, 
\label{Vpp_def}
\end{equation}
subtracts from the expression for the self-volume of atom $j$ all those overlap volumes involving both atoms $i$ and $j$.
Two differences with respect to the original AGBNP1 formulation are
introduced. The first is that $s_j$ [Eq.\ (\ref{sj_def})] is computed
from the self-volume after subtracting from it the volume of the
region subtended by the solvent-exposed surface area in between the
enlarged and van der Waals atomic spheres of atom $j$. Referring to
Fig.\ \ref{surfvol.fig} the volume of this region is $d_j A_j$ as in
Eq.\ (\ref{sj_def}) with
\begin{equation}
d_j = \frac13 R'_j \left[ 1 - \left( \frac{R_j}{R'_j} \right)^3
\right] \, .
\label{surfvol}
\end{equation}
The other difference concerns the $V'_{ji}$ term which in the AGBNP1
formulation is approximated by the
2-body overlap volume $V_{ij}$; the first term in the right hand side of Eq.\
(\ref{Vpp_def}). This approximation is found to lack  sufficient
accuracy for the the present formulation given the relative increase
in size of all overlap volumes. In the present formulation $V'_{ji}$
is computed at all overlap orders according to Eq.\ (\ref{Vpp_def}).

In the AGBNP2 formulation the functional form for the pair
descreening function $Q_{ji}$ is the same as in the original formulation
(see Appendix of reference \cite{Gallicchio:Levy:2004}), however in the new formulation this
function is evaluated using the van der Waals radius $R_i$ for atom
$i$ (the atom being ``descreened'') and the augmented radius $R^{\rm c}_j$
for atom $j$ (the atom that provides the solvent descreening), rather than using the van
der Waals radius for both atoms. So if the pair
descreening function is denoted by $Q(r, R_1, R_2)$, where $r$ is the
interatomic distance, $R_1$ the radius of the atom being descreened
and $R_2$ the radius that provides descreening, we set in Eq.\
(\ref{Bpairwise_sji})
\begin{equation}
Q_{ji} = Q(r_{ij},R_i,R^{\rm c}_j) \, .
\end{equation} 

%It also avoids the separate calculation of the cavity energy
%since the same volume description is applied to both the GB and van
%der Waals energies and the cavity energy.

\section{Hydrogen Bonding Corrections}

In this section we present the analytical model that implements the
short--range hydrogen-bonding correction function for AGBNP2.
The model is based on a geometrical procedure to measure the
degree to which a solute atom can interact with hydration sites on the
solute surface. The procedure is as follows. A sphere of radius
$R_{s}$ representing a water molecule is placed in a position
that provides near optimal interaction with a hydrogen bonding donor
or acceptor atom of the solute. The position $\mathbf{r}_s$ of this
water sphere $s$ is function of the positions of
two or more parent atoms that compose the functional group including
the acceptor/donor atom:
\begin{equation}
\mathbf{r}_s = \mathbf{r}_s(\left\{ \mathbf{r}_{ps} \right\})
\end{equation}
where $\left\{ \mathbf{r}_{ps} \right\}$ represents the positions of
the set of parent atoms of the water site $s$. For instance the water site
position in correspondence with a polar hydrogen is:
\[
\mathbf{r}_s = \mathbf{r}_D + 
\frac{\mathbf{r}_H - \mathbf{r}_D}{\left| \mathbf{r}_H - \mathbf{r}_D \right|} d_{\rm HB}
\]
where $\mathbf{r}_D$ is the position of the heavy atom donor,
$\mathbf{r}_H$ is the position of the polar hydrogen, and $d_{\rm HB}$
is the distance between the heavy atom donor and the center of the
water sphere (see Fig. \ref{ws.fig}). Similar relationships (see
Appendix) are employed to place candidate water spheres in
correspondence with hydrogen bonding acceptor atoms of the
solute. These relationships are based on the local topology of the
hydrogen bonding acceptor group (linear, trigonal, and tetrahedral).
This scheme places one or two water spheres in correspondence
with each hydrogen bonding acceptor atom (see Table \ref{tab.hs}).

\begin{figure}[tbp]
\begin{center}
\includegraphics{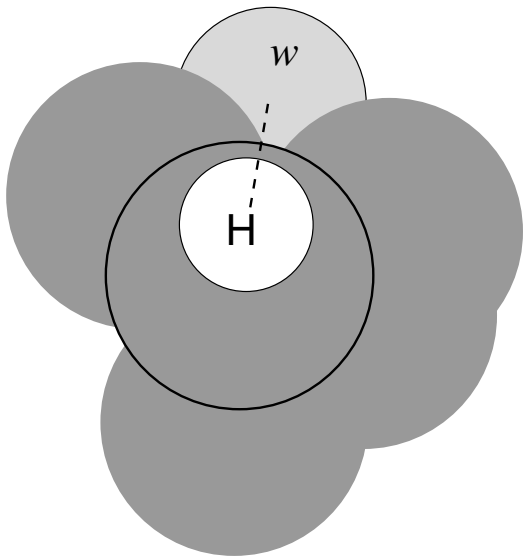}
\end{center}
\caption{\label{ws.fig} {\small Schematic diagram for the placement of the
  water sphere ($w$, light gray) corresponding to the hydrogen bonding
  position relative to the a polar hydrogen (white sphere) of the
  solute (dark gray).  The dashed line traces the direction of the
  hydrogen-parent heavy atom (circled) bond along which the water
  sphere is placed. The magnitude of hydrogen bonding correction grows
  as a function of the volume (light gray) of the water site sphere
  not occupied by solute atoms.}}
\end{figure}

The magnitude of the hydrogen bonding correction corresponding to each
water sphere is a function of the predicted water occupancy of the
location corresponding to the water sphere. In this work the water occupancy 
is measured by the fraction $w_s$ of the volume of the water site sphere
that is accessible to water molecules without causing steric clashes
with solute atoms (see Fig.\ (\ref{ws.fig})
\begin{equation}
w_s = \frac{V_s^{\rm free}}{V_s}
\label{s_occupancy}
\end{equation}
where $V_s = (4/3) \pi R_s^3$ is the volume of the water sphere
and
\begin{equation}
V_s^{\rm free} = V_s - \sum_i V_{si} + \sum_{i<j} V_{sij} -
\sum_{i<j<k} V_{sijk}
\label{s_free_volume}
\end{equation}
is the ``free'' volume of water site $s$, obtained by summing over the
two-body, three-body, etc.\ overlap volumes of the water sphere with
the solute atoms. Note that the expression of the free volume is the
same as the expression for the self-volume except that it lacks the fractional
coefficients $1/2$, $1/3$, etc. The overlap volumes in Eq.\
(\ref{s_free_volume}) are computed using radius $R_s$ for the water
site sphere (here set to $1.4$ \AA) and augmented radii $R^{\rm c}_i$ for the
solute atoms. Eq.\
(\ref{s_free_volume}) is derived similarly to the expression for the
self volumes by removing overlap volumes from the volume of the water
sphere rather than evenly distributing them across the atoms
participating in the overlap.

\begin{figure}[tbp]
\begin{center}
\scalebox{0.75}{\includegraphics{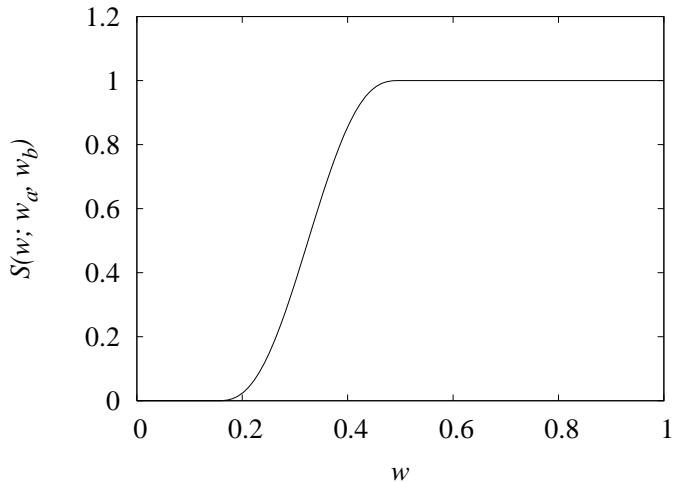}}
\end{center}
\caption{\label{switch.fig} {\small The switching function $S(w; w_a, w_b)$
  from Eqs.\ (\ref{switch1}) and (\ref{switch2}) with $w_a = 0.5$ and
  $w_b = 0.9$. }
}
\end{figure}

Given the water occupancy $w_s$ of each water sphere, the expression for the
hydrogen bonding correction for the solute is
\begin{equation}
\Delta G_{\rm HB} = \sum_s h_s S(w_s; w_a, w_b) \, ,
\label{HB_corr_energy}
\end{equation}
where $h_s$ is the maximum correction energy which depends on the type
of solute-water hydrogen bond (see Table \ref{tab.hs}), and $S(w; w_a,
w_b)$ is a polynomial switching function which is zero for $w < w_a$,
one for $w > w_b$, and smoothly (with continuous first derivatives)
interpolates from 0 to 1 between $w_a$ and $w_b$ (see Fig.\
\ref{switch.fig}). The expression of $S(w; w_a, w_b)$ is:
\begin{equation}
S(w; w_a, w_b) = \left\{ \begin{array}{ll}
0 & w \le w_a \\
f_w\left( \frac{w-w_a}{w_b - w_a}\right) & w_a < w < w_b  \\
1 & w \ge w_b
\end{array} \right.
\label{switch1}
\end{equation}
where $f_w(x)$ is a switching function given by Eq.\ (\ref{switch2}).
In this work we set $w_a = 0.15$ and $w_b = 0.5$. This scheme
establishes (see Fig.\ \ref{switch.fig}) that no correction is applied
if more than 75\% of the water sphere volume is not water accessible,
whereas maximum correction is applied if 50\% or more of the water
sphere volume is accessible.

\begin{table}[tbp]
\renewcommand{\baselinestretch}{1.0}\small\normalsize
\caption{\label{tab.hs} Optimized surface tension parameters and
  hydrogen bonding correction parameters for the
  atom types present in protein molecules. $\gamma$ is the surface
  tension parameter, $N_w$ is the number of
  water spheres and $h$ is the maximum correction corresponding to
  each atom type [Eq.\ (\ref{HB_corr_energy})]. Atom types not listed
  do not have HB corrections and 
  are assigned $\gamma = 117$ cal/mol/\AA$^2$}
\begin{center}
\begin{tabular}{lcccc} \hline
Atom Type & $\gamma$ (cal/mol/\AA$^2$) & Geometry & $N_w$ & $h$ (kcal/mol) \\ \hline
C (aliphatic)  & 129 & & & \\
C (aromatic)   & 120 & & & \\
H on N          &      & linear      & 1   & -0.25 \\
H on N (Arg)    &      & linear      & 1   & -2.50 \\
H on O          &      & linear      & 1   & -0.40 \\
H on S          &      & linear      & 1   & -0.50 \\
O (alchol)      &117      & tetrahedral & 2   & -0.40 \\
O (carbonyl)    &117      & trigonal    & 2   & -1.25 \\
O (carboxylate) &40       & trigonal    & 2   & -1.80 \\
N (amine)       &117      & tetrahedral & 1   & -2.00 \\
N (aromatic)    &117      & trigonal    & 1   & -2.00 \\
S               &117      & tetrahedral & 2   & -0.50
\end{tabular}
\end{center}
\renewcommand{\baselinestretch}{1.5}\small\normalsize
\end{table}

\subsection{Molecular dynamics of mini-proteins}
\label{md.sect}

To be written (usual MD details, time-step, etc.).

\section{Results}

\subsection{Accuracy of Born Radii and Surface Areas}

The quality of any implicit solvent model depends primarily on the
reliability of the physical model on which it is based. The accuracy
of the implementation, however, is also a critical aspect for the
success of the model in practice. This is true in particular for
models, such as AGBNP, based on the Generalized Born formula. It has
been pointed out, for instance, that approximations in the integration
procedure to obtain the Born radii may actually be of more
significance than the physical approximations on which the GB model is
based.\cite{Onufriev:Case:Bashford:2002} It is therefore important to
test that the conformational dependent quantities employed by AGBNP2
are a good approximation to the geometrical parameters that they are
supposed to represent.  The present AGBNP2 formulation relies mainly on
three types of conformational dependent quantities: Born radii [Eq.\
(\ref{Bpairwise_sji})], solvent accessible surface areas (\ref{Aidef}), and
solvent accessibilities of hydration sites [Eq.\ (\ref{switch1})]. In
this section we analyze the validity of the AGBNP2 analytical estimates
for the Born radii and surface areas against accurate numerical
results for the same quantities.   

\begin{figure}[tbp]
\begin{center}
\scalebox{0.9}{\includegraphics{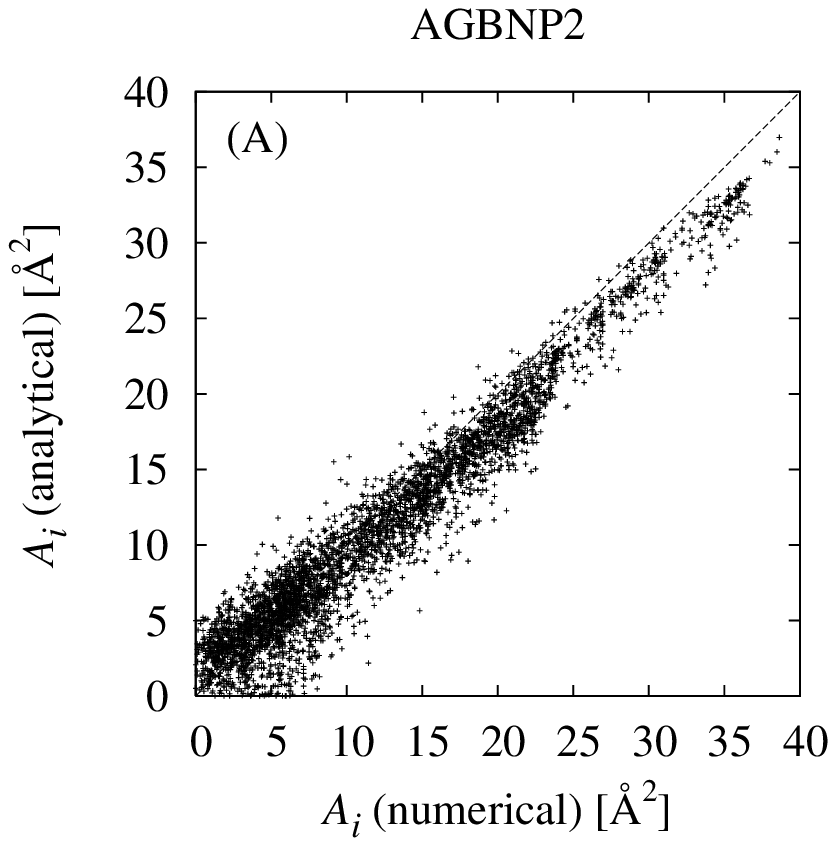}%
\hspace{-1.5in}\includegraphics{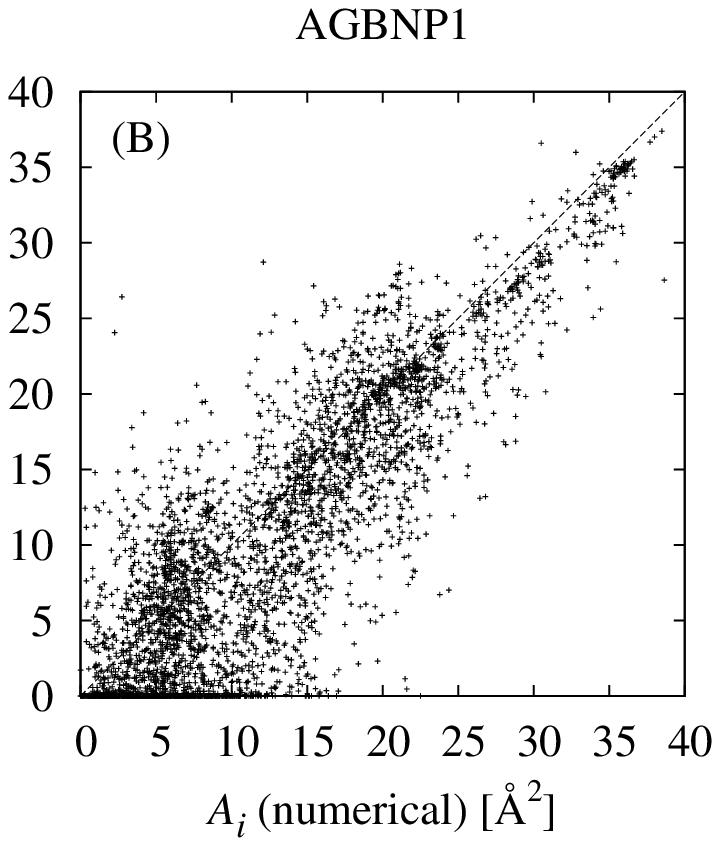}}
\end{center}
\caption{\label{surf.fig} {\small Comparisons between numerical and
  analytical molecular surface areas of the heavy atoms of the crystal 
  structures (1ctf and 1lz1, respectively) of the C-terminal domain of the ribosomal protein
  L7/L12 (74 aa) and human lysozyme (130 aa), and of four
  conformations each of the trp-cage, cdp-1, and fsd-1 miniproteins
  extracted from the corresponding explicit solvent MD
  trajectories of the same
  protein conformations as in Fig.\ \ref{br.fig}. (A) Analytical
  molecular surface areas computed using the present model, and (B),
  for comparison, analytical surface areas computed using the original
  model from reference \cite{Gallicchio:Levy:2004}.
}
}
\end{figure}

We employ the GEPOL program\cite{Pascual:Silla:90} to compute numerically atomic
solvent accessible surface areas with a solvent probe diameter of 1
\AA, the same probe diameter that defines the solute-solvent boundary
in the AGBNP model. Fig.\ \ref{surf.fig}A shows the comparison between
the surface area estimates given by the present formulation of AGBNP
and the numerical surface areas produced by GEPOL for a series of
native and modeled protein conformations. In Fig.\
\ref{surf.fig}B we show the same comparison for the surface areas of
the original AGBNP1 model. These representative results
show that the present analytical surface area implementation, which as
described above employs a weaker switching function for the overlap
volumes, produces significantly more accurate atomic surface areas
than the original model. These improvements in the computation of the
surface ares, introduced mainly to obtain more accurate Born radii
through Eq.~(\ref{sj_def}), are also expected to yield more reliable
cavity hydration free energies differences.

\begin{figure}[tbp]
\begin{center}
\scalebox{0.9}{\includegraphics{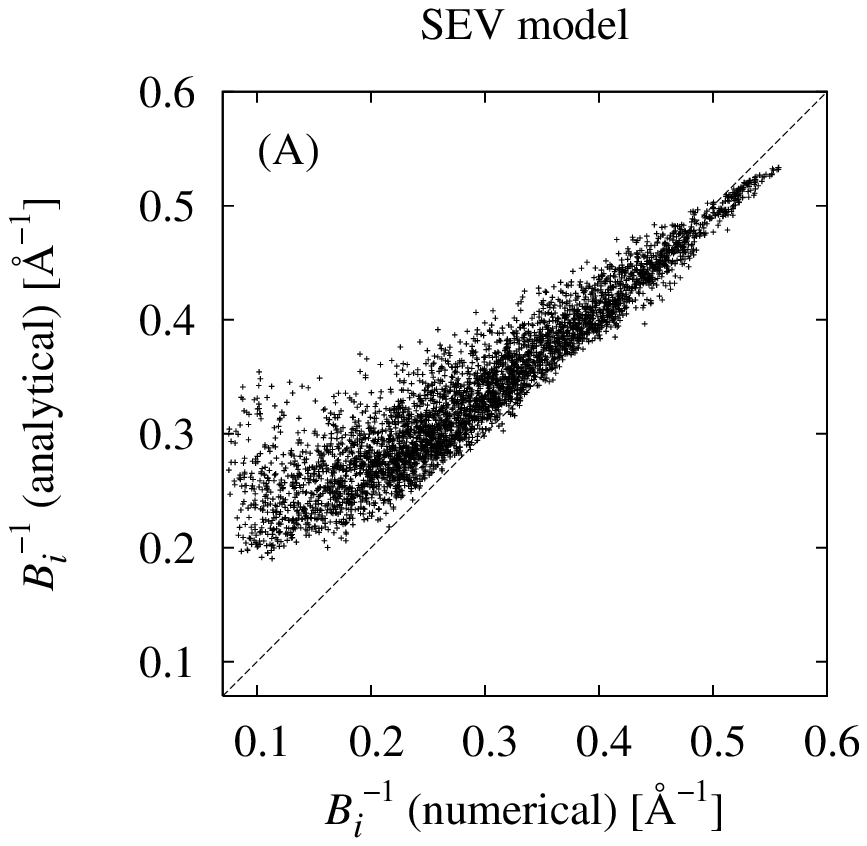}\hspace{-1.5in}\includegraphics{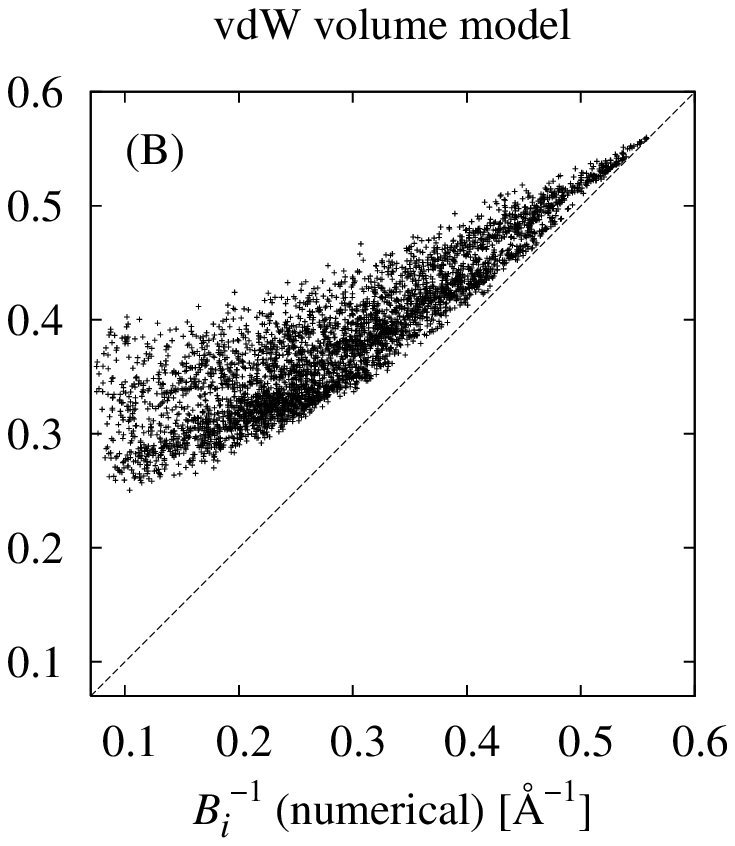}}
\end{center}
\caption{\label{br.fig} {\small Comparisons between numerical and
  analytical inverse Born radii for the heavy atoms of the same
  protein conformations as in Fig.\ \ref{surf.fig}
(A) Analytical Born radii computed using the present
  SEV model. (B) Analytical Born radii computed using the van der
  Waals volume model (reference \cite{Gallicchio:Levy:2004}).   }
}
\end{figure}

Fig.~\ref{br.fig} illustrates on the same set of protein conformations
the accuracy of the inverse Born radii, $B_i^{-1}$, obtained using the AGBNP2
pairwise descreening method using the SEV model for the solute volume
described above (Eq.~(\ref{Bpairwise_sji})), by comparing them to
accurate estimates obtained by evaluating the integral in
Eq.~(\ref{Bintegral}) numerically on a grid. The comparison is
performed for the inverse Born radii because these, being proportional
to GB self-energies, are a more reliable indicators of accuracy than
the Born radii themselves. The grid for the numerical integration was
prepared as previously reported,\cite{Gallicchio:Levy:2004} except that the solvent
excluding volume (SEV) of the solute was employed here rather than the
van der Waals volume. The integration grid over the SEV
was obtained by taking advantage of the particular way that the GEPOL
algorithm describes the SEV of the solute; GEPOL
iteratively places auxiliary spheres of various dimensions in the
interstitial spaces between
solute atoms in such a way that the van der Waals volume of the solute
plus the auxiliary spheres accurately reproduces the SEV of the
solute. Therefore in the present application a grid point was
considered part of the SEV of the solute if it was contained within
any solute atom or any one of the auxiliary spheres placed by
GEPOL. The default 1.4 \AA\ solvent probe radius was chosen for the
numerical computation of the SEV with the GEPOL program. 

The results
of this validation (Fig.~\ref{br.fig}) show that the analytical SEV
pairwise descreening model described above is able to yield Born radii
which are not as affected by the spurious high dielectric
interstitial spaces present in the van der Waals volume description of
the solute. 
With the van der Waals volume model (Fig.~\ref{br.fig}B)
the Born radii of the majority of solute atoms start to significantly
deviate from the reference  values for Born radii
larger than about 2.5 \AA\ ($B^{-1} = 0.4$ \AA$^{-1}$). Born radii
computed with the analytical SEV model instead (Fig.~\ref{br.fig}A) track the
reference values reasonably well further up to about 4 \AA\ ($B^{-1} = 0.25$
\AA$^{-1}$). Despite this significant improvement most Born radii are
still underestimated by the improved model (and, consequently, the
inverse Born radii are overestimated - see Fig.~\ref{br.fig}), particularly those of
non-polar atoms near the hydrophobic core of the larger
proteins. These regions tend to be loosely packed and tend to contain
interstitial spaces too large to be correctly handled by the present
model. Because it mainly involves groups of low polarity this
limitation has a small effect on the GB solvation energies. It has
however a more significant effect on the van der Waals solute--solvent interaction
energy estimates through Eq.~(\ref{Gvdw_def}), which systematically
overestimate the magnitude of the interaction for atoms buried in
hydrophobic protein core. While the present model in general meliorates in all
respects the original AGBNP model, we are currently exploring ways to address
this residual source of inaccuracy. 

\subsection{Small Molecule Hydration Free Energies}

The validation and parametrization of the hydrogen bonding and cavity
correction parameters have been performed based
on the agreement between experimental and predicted AGBNP2 hydration
free energies of a selected set of small molecules, listed in Table
\ref{tab.small.mol}, containing the main functional groups present in
proteins.  This set of molecules includes only small and relatively
rigid molecules whose hydration free energies can be reliably
estimated using a single low energy representative
conformation\cite{Mobley:Chodera:Dill:2008} as it was done here. Table
\ref{tab.small.mol} lists for each molecule the experimental hydration
free energy, the AGBNP2 hydration free energy computed
without HB corrections and the
default $\gamma = 117$ cal/mol/\AA$^2$ surface tension
parameter, denoted by AGBNP2/SEV, as well as the hydration free energy from the
AGBNP2 model including the HB
correction term and the parameters listed in Table \ref{tab.hs}.

\begin{table}[tbp]
\renewcommand{\baselinestretch}{1.0}\small\normalsize
\caption{\label{tab.small.mol} 
  Experimental and predicted hydration free energies of a set of small
  molecules.
}
\begin{center}
\begin{tabular}{lccc} \hline
%Molecule       & Exp.$^{a,b}$ & AGBNP (no corr.)$^{a,c}$  & AGBNP (with corr.)$^{a,d}$\\ \hline
Molecule       & Exper. & AGBNP/SEV  & AGBNP2\\ \hline
n-ethane       &   1.83 &   0.98 &   1.80  \\
n-propane      &   1.96 &   0.92 &   1.97  \\
n-butane       &   2.08 &   0.88 &   2.14  \\
n-pentane      &   2.33 &   0.78 &   2.26  \\
n-hexane       &   2.50 &   0.70 &   2.40  \\
cyclo-pentane  &   1.20 &   0.34 &   1.63  \\
cyclo-hexane   &   1.23 &   0.05 &   1.50  \\ 

benzene        &  -0.87 &  -1.50 &  -1.14  \\      
toluene        &  -0.89 &  -1.66 &  -0.94  \\ \hline

acetone        &  -3.85 &  -1.09 &  -3.83  \\
acetophenone   &  -4.58 &  -2.74 &  -5.07  \\ \hline

ethanol        &  -5.01 &  -4.77 &  -5.30  \\
phenol         &  -6.62 &  -4.51 &  -5.38  \\
ethandiol      &  -9.60 &  -7.99 &  -9.87  \\ \hline

acetic acid    &  -6.70 &  -2.73 &  -7.05  \\
propionic acid &  -6.48 &  -2.58 &  -6.38  \\ \hline

methyl acetate &  -3.32 &  -0.10 &  -3.92  \\ 
ethyl acetate  &  -3.10 &  -0.02 &  -3.60  \\ \hline

methyl amine   &  -4.56 &  -2.39 &  -4.37  \\
ethyl amine    &  -4.50 &  -2.24 &  -3.95  \\
dimethyl amine &  -4.29 &  -1.95 &  -3.21  \\ 
trimethyl amine&  -3.24 &  -1.78 &  -2.39  \\ \hline

acetamide        &  -9.71 & -6.81 & -10.45 \\
N-methylacetamide& -10.08 & -4.75 &  -7.51 \\ \hline

pyridine         &  -4.70 &  -3.62 & -5.30 \\ 
2-methylpyridine &  -4.63 &  -2.94 & -4.22 \\
3-methylpyridine &  -4.77 &  -2.82 & -4.13 \\ \hline

methanethiol & -1.24 & -0.61 & -1.46 \\
ethanethiol  & -1.30 & -0.57 & -1.22 \\ \hline

acetate ion        & -79.90 & -77.32 & -87.70 \\
propionate ion     & -79.10 & -76.29 & -86.29 \\
methylammonium ion & -71.30 & -73.21 & -73.54 \\
ethylammonium ion  & -68.40 & -70.63 & -70.75 \\ 
methylguanidinium   & -62.02$^e$ & -57.30 & -69.81 \\ \hline 
\end{tabular}
\end{center}
$^a$ In kcal/mol.\newline
$^b$ Experimental hydration free energy from reference
\cite{Cabani:Gianni:Mollica:Lepori:81} except where indicated.\newline
$^c$ AGBNP predicted hydration free energies with the default $\gamma$
parameter for all atoms types ($\gamma = 117$ cal/mol/\AA$^2$) and
without HB corrections.\newline
$^d$ AGBNP predicted hydration free energies with optimized parameters
listed in Table \ref{tab.hs}.
$^e$ From reference \cite{Vorobyov:Li:Allen:2008}.

\renewcommand{\baselinestretch}{1.5}\small\normalsize
\end{table}

Going down the results in Table \ref{tab.small.mol} we notice a
number of issues addressed by the new implementation. With the new
surface area implementation and without corrections (third column in
Table \ref{tab.small.mol}) the hydration free energies of the normal
alkanes are too small compared to experiments, furthermore, in
contrast with the experimental behavior, the predicted hydration free
energies incorrectly become more favorable with increasing chain
length. A similar behavior is observed for the aromatic
hydrocarbons. Clearly this is due to the rate of increase of the
positive cavity term with increasing alkane size which is insufficient
to offset the solute--solvent van der Waals interaction energy term,
which becomes more negative with increasing solute size. We have
chosen to address this shortcoming by increasing by 10.2\% and 2.5\%
respectively the effective surface tensions for aliphatic and aromatic
carbon atoms rather than decreasing the corresponding $\alpha$
parameters of the van der Waals term since the latter had been
previously validated against explicit solvent simulations. We have
chosen to limit the increases of the surface tension parameters to
aliphatic and aromatic carbon atoms since the results for polar
functional groups did not indicate that this change was necessary for the
remaining atom types. With this new parametrization we achieve
(compare the second and fourth column in Table \ref{tab.small.mol})
excellent agreement between the experimental and predicted hydration
free energies of the alkanes and aromatic compounds. Note that the
AGBNP2 model, regardless of the parametrization, correctly predicts the more
favorable hydration free energies of the cyclic alkanes relative to
their linear analogous. AGBNP2, thanks to its unique decomposition of
the non-polar solvation free energy into unfavorable cavity term and
an opposing favorable term, is, to our knowledge, the only analytic
implicit solvent implementation capable of describing correctly this
feature of the thermodynamics of hydration of hydrophobic solutes. 

The AGBNP2 model without corrections markedly underpredicts the
magnitudes of the experimental hydration free energies of the
compounds containing carbonyl groups (ketones, organic acids, and
esters). The hydration free energies of alcohols are also
underpredicted but by
smaller amounts. Much better agreement with the experimental hydration
free energies of these oxygen-containing compounds (see Table
\ref{tab.small.mol}) is achieved after applying hydrogen bonding
corrections with $h = -1.25$ kcal/mol for the carbonyl oxygen, and $h =
-0.4$ kcal/mol for both the hydrogen and oxygen atoms of the hydroxy
group (Table \ref{tab.hs}). Note that the same parameter employed
individually for carbonyl and hydroxy groups in ketones and alcohols
are applied to the more complex carboxylic groups of acids and esters
as well as amides and carboxylate ions. The thiol group of organic
sulfides required similar corrections as the hydroxy group (Tables
\ref{tab.hs} and \ref{tab.small.mol}). The AGBNP2 model without 
corrections also markedly underpredicted the magnitude of the
experimental hydration free energies of amines and amides, and, to a
smaller extent, of compounds with nitrogen containing etherocyclic aromatic
rings. The addition of HB corrections of $-0.25$ kcal/mol for amine
hydrogens and $h=-2.0$ kcal/mol for both amine and aromatic nitrogen
atoms yields improvement agreement (Table \ref{tab.small.mol}),
although the effect of N-methylation is still over-emphasized by both
AGBNP2 parametrization.

\subsection{Mini-protein results}

As described in Section \ref{md.sect} we have performed restricted MD
simulations of a series of so-called mini-proteins (trp-cage, cdp-1,
and fsd-1) to study the extent of the agreement between the
conformational ensembles generated with the original AGBNP
implementation (AGBNP1) and the present implementation (AGBNP2) with
respect to explicit solvent-generated ensembles. Earlier
studies suggested that the AGBNP/OPLS-AA model correctly
reproduced for the most part backbone secondary structure features of
protein and peptides. The tests in the present study are therefore
focused on sidechain conformations. The backbone atoms were
harmonically restricted to remain within approximately 3 \AA\
C-$\alpha$ RMSD of the corresponding NMR experimental models. We
structurally analyzed the ensembles in terms of the extent of
occurrence of intramolecular interactions.  

As shown in Table \ref{hbt.table} we measured a significantly higher
average number of intramolecular hydrogen bonds and ion pairing in the
AGBNP1 ensembles relative to the explicit solvent ensembles for all
mini-proteins studied. The largest deviations are observed for cdp-1
and fsd-1, two mini-proteins particularly rich in charged sidechains,
with on average nearly twice as many intramolecular hydrogen bonds
compared to explicit solvent. Many of the excess intramolecular
hydrogen bonds with AGBNP1 involve interactions between polar groups
(polar sidechains or the peptide backbone) and the sidechains of
charged residues. For example for cdp-1 we observe (see Table
\ref{hbt.table}) approximately 8 hydrogen bonds between polar and
charged groups on average compared to nearly none with explicit
solvation. 

Despite the introduction of empirical surface area corrections to
penalize ion pairs,\cite{Felts:Gallicchio:Levy:2008} AGBNP1 over-predicts ion-pair
formation. We found that ion pairing involving arginine were
particularly over-stabilized by AGBNP1 as we observed stable ion
pairing between arginine and either glutamate or aspartate residue
during almost the entire duration of the simulation in virtually all
cases in which this was topologically feasible given the imposed
backbone restrains. In contrast, with explicit solvation some of the
same ion pairs were seen to form and break numerous times indicating a
balanced equilibrium between contact and solvent-separated
conformations. This balance is not reproduced with implicit solvation
which instead strongly favors ion pairing. In any case, the relative
stability of ion pairs appeared to depend in subtle ways on the
protein environment as for example the two ion pairs between arginine
and glutamate of cdp-1 were found to be stable with either explicit
solvation or AGBNP1 implicit solvation whereas other Arg-Glu ion pairs
in trp-cage and fsd-1 were found to be stable only with implicit
solvation.

\begin{table}[tbp]
\renewcommand{\baselinestretch}{1.0}\small\normalsize
\caption{\label{hbt.table} Average number of some types of
intramolecular electrostatic interactions in the explicit solvent
conformational ensembles, and the AGBNP1 and AGBNP2 ensembles of the
trp-cage, cdp-1, and fsd-1 mini-proteins.  }
\begin{center}
\begin{tabular}{lccc} \hline
Mini-protein  & Explicit & AGBNP1 & AGBNP2 \\ \hline
\multicolumn{4}{c}{Intramolecular Hydrogen Bonds} \\
trp-cage      & 13.5    & 18.3	& 15.3  \\
cdp-1         & 12.6    & 24.5	& 15.4  \\
fsd-1         & 14.1    & 24.6    & 14.3  \\
\multicolumn{4}{c}{Polar--Polar Hydrogen Bonds} \\
trp-cage      & 12.9    & 17.1	& 13.8  \\
cdp-1         & 12.5    & 16.4	& 14.1  \\
fsd-1         & 12.0    & 14.2    & 12.9  \\
\multicolumn{4}{c}{Polar--Charged Hydrogen Bonds} \\
trp-cage      & 0.6     & 1.2	& 1.4  \\
cdp-1         & 0.1     & 8.1	& 1.3  \\
fsd-1         & 2.1     & 9.6     & 1.4  \\
\multicolumn{4}{c}{Ion Pairs} \\
trp-cage      & 0.3    & 1.0	& 1.0  \\
cdp-1         & 2.5    & 2.9	& 2.7  \\
fsd-1         & 1.4    & 4.6      & 4.0  \\
\end{tabular}
\end{center}
\renewcommand{\baselinestretch}{1.5}\small\normalsize
\end{table}

\begin{figure}[tbp]
\begin{center}
\scalebox{0.75}{\includegraphics{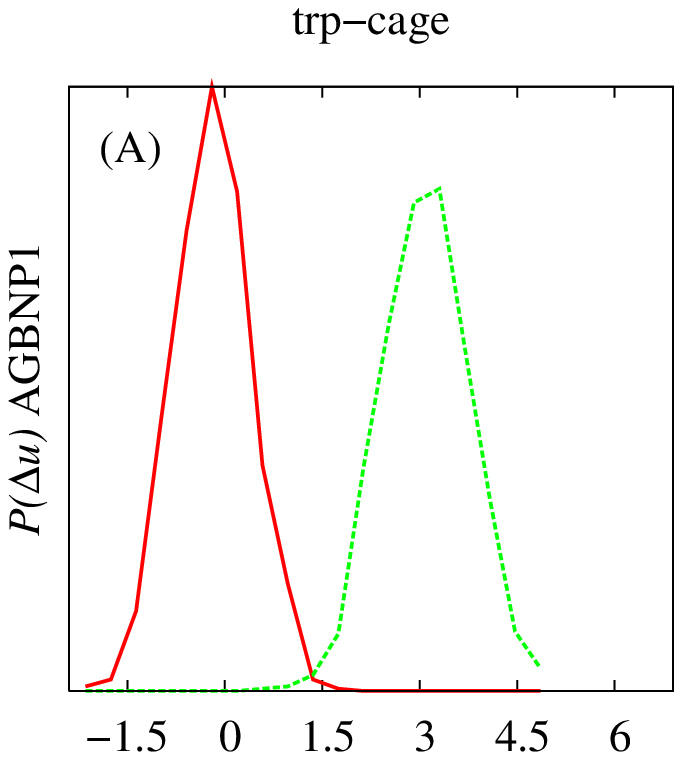}\hspace{-2.45in}\includegraphics{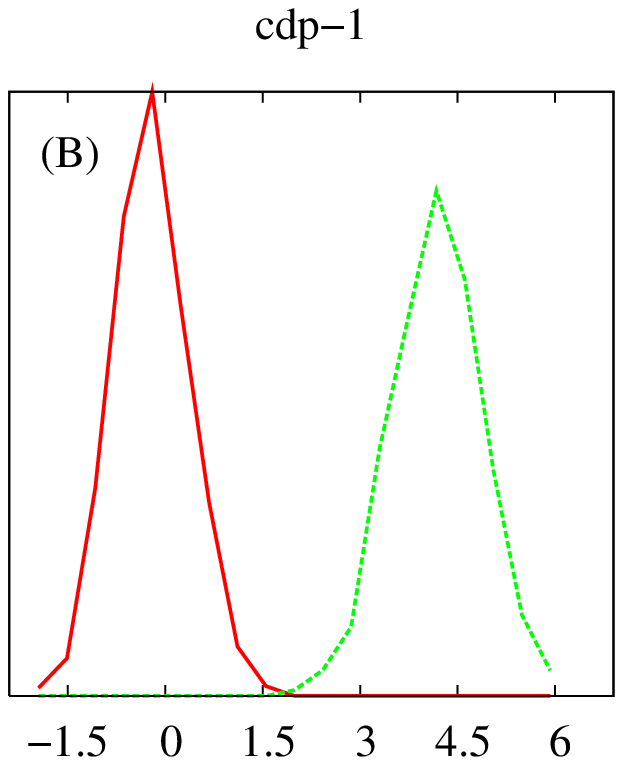}\hspace{-2.5in}\includegraphics{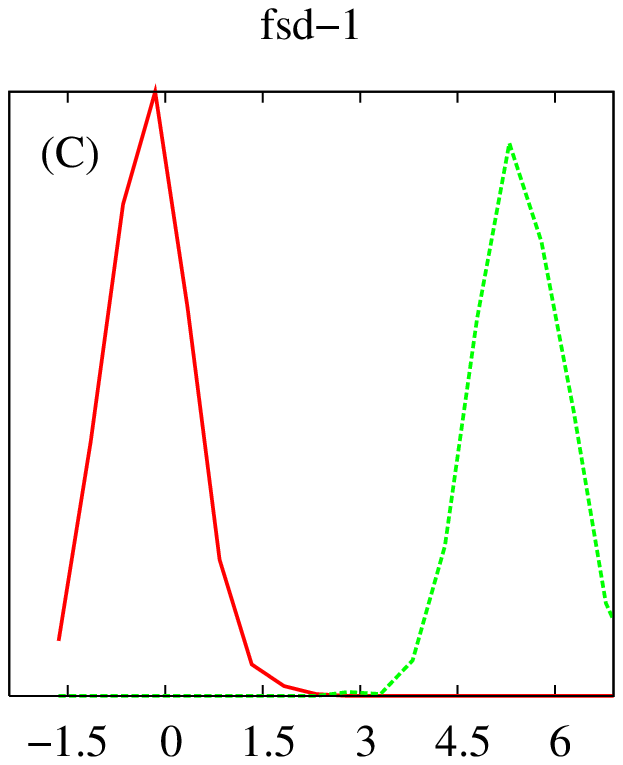}}\\
\vspace{-0.6in}\noindent
\scalebox{0.75}{\includegraphics{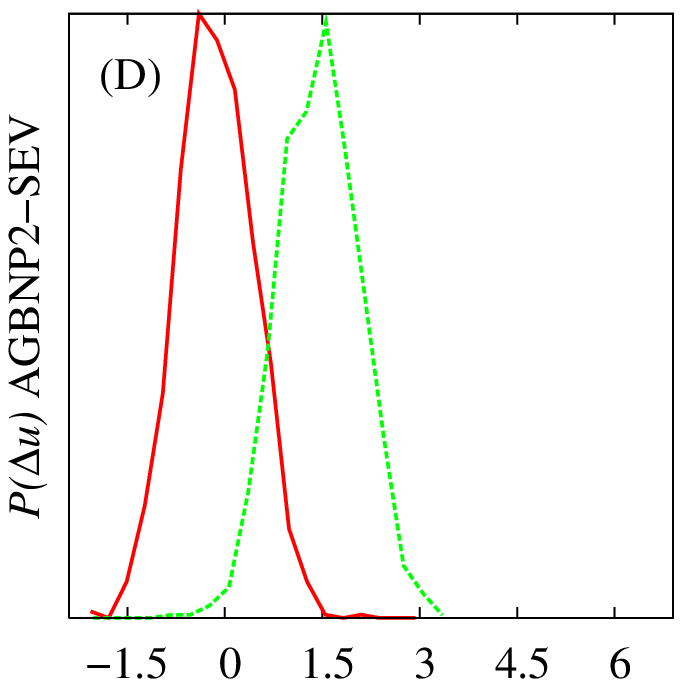}\hspace{-2.45in}\includegraphics{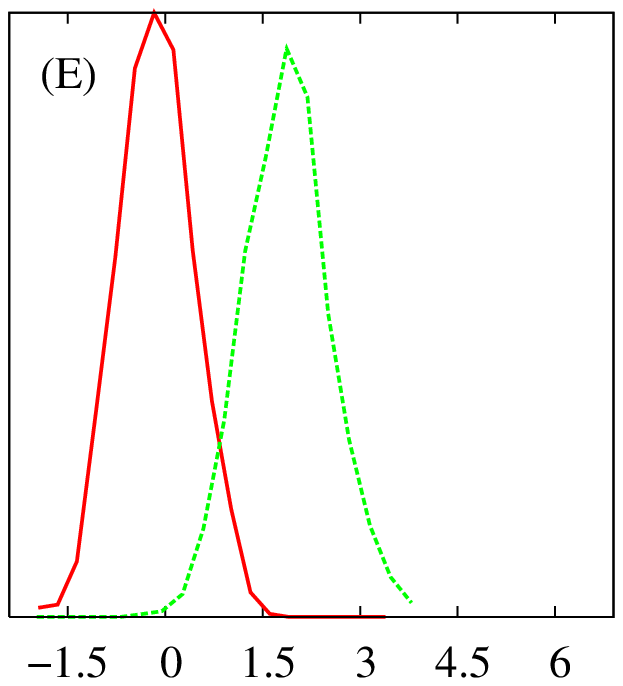}\hspace{-2.5in}\includegraphics{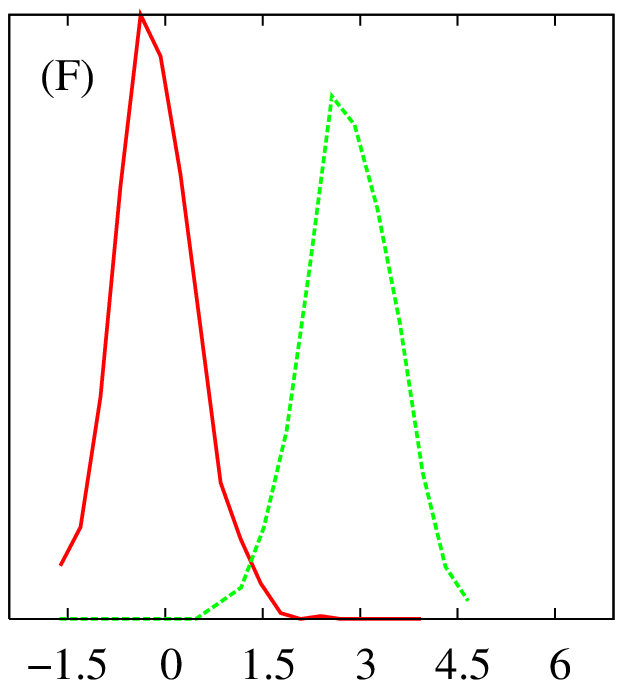}}\\
\vspace{-0.6in}\noindent
\scalebox{0.75}{\includegraphics{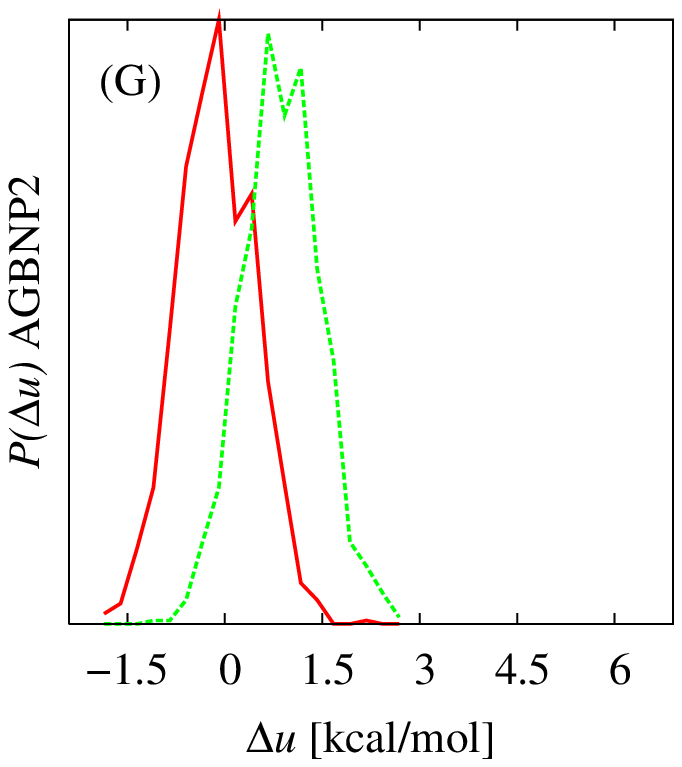}\hspace{-2.45in}\includegraphics{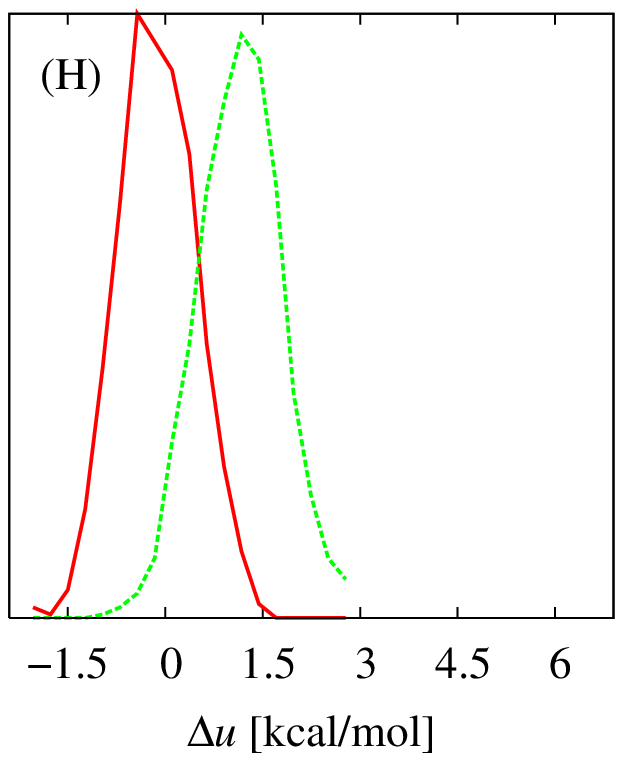}\hspace{-2.5in}\includegraphics{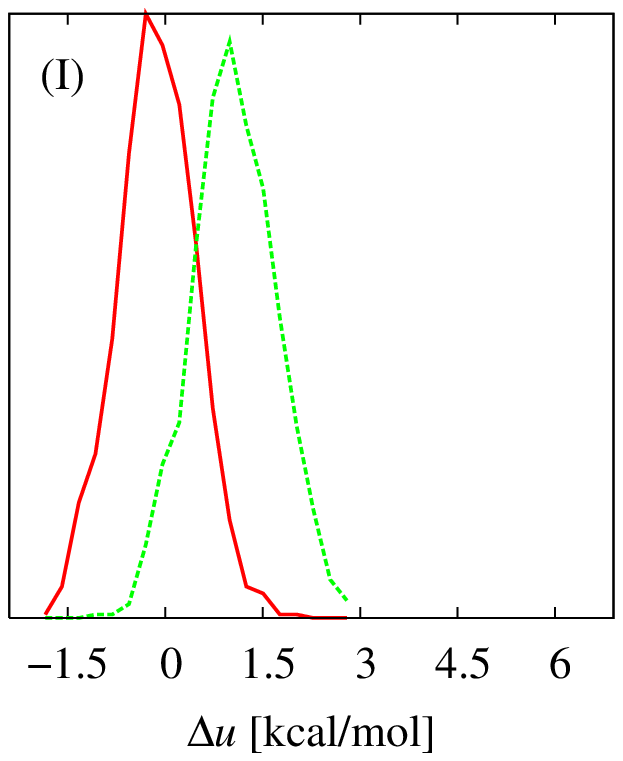}}
\end{center}
\caption{\label{agbnp1de.fig} {\small Probability distributions of the
AGBNP1/OPLS-AA (first row, panels A, B, and C), AGBNP2-SEV/OPLS-AA
(second row, panels D, E, and F), and AGBNP2/OPLS-AA effective
potential energies for the conformational ensembles for the trp-cage
(first column, panels A, D, G), cdp-1 (second column, panels B, E, H),
and fsd-1 (third column, panels C, F, I) mini-proteins obtained using
the AGBNP1 solvation model (full line) and explicit solvation
(dashed line). The AGBNP2-SEV model is the AGBNP2 model with the SEV
representation of the solute volume and without the hydrogen bonding
corrections [$h_s=0$ in Eq.~(\ref{HB_corr_energy}) for all atom
types].  The distributions are shown as a function of the energy gap
per residue ($\Delta u$) relative to the mean effective potential
energy of the implicit solvent ensemble.  } }
\end{figure}

This analysis generally confirms quantitatively a series of past
observations made in our laboratory indicating that the original AGBNP
implementation tends to be biased towards conformations with excessive
intramolecular electrostatic interactions, at the expense of more
hydrated conformations in which polar groups are oriented so as to
interact with the water solvent. During the process of development of
the modifications to address these problems, we found useful to
rescore with varying AGBNP2 formulations and parametrization the
mini-protein conformational ensembles obtained with AGBNP1 and
explicit solvation, bypassing the time consuming MD runs to obtain
ensembles with each new parametrization. An example of this analysis
is shown in Fig.~\ref{agbnp1de.fig}, which compares the probability
distributions of the AGBNP1 and AGBNP2 effective potential energies
over the conformational ensembles generated with AGBNP1 implicit
solvation and with explicit solvation.  These results clearly show
that the AGBNP1/OPLS-AA effective potential disfavor conformations
from the explicit solvent ensemble relatively to those generated with
implicit solvation. The AGBNP1 energy scores (first row of
Fig.~\ref{agbnp1de.fig}) of the explicit solvent ensembles of all
mini-proteins are shifted towards higher energies than those of the
AGBNP1 ensemble, indicating that conformations present in the explicit
solvent ensemble would be rarely visited when performing
conformational sampling with the AGBNP1/OPLS-AA
potential. AGBNP1/OPLS-AA assigns a substantial energetic penalty (see
Fig.~\ref{agbnp1de.fig} A--C) to the explicit solvent ensemble
relative to the AGBNP1 ensemble (on average $3.3$, $4.4$, and $5.7$
kcal/mol per residue for, respectively, the trp-cage, cdp-1, and fsd-1
mini-proteins). This energetic penalty, being significantly larger than thermal
energy, rules out the possibility that conformational entropy effects
could offset it to such an extent so as to equalize the relative free
energies of the two ensembles. Detailed analysis of the energy scores
shows that, as expected, the AGBNP1 implicit solvent ensemble is
mainly favored by more favorable electrostatic Coulomb interaction
energies due to its greater number of intramolecular electrostatic
contacts relative to the explicit solvent ensemble (see
above). Conversely, the AGBNP1 solvation model does not assign
sufficiently favorable hydration free energy to the more
solvent-exposed conformations obtained in explicit solvent so as to
make them competitive with the more compact conformations of the
AGBNP1 ensemble.

The energy distributions of the AGBNP2/OPLS-AA effective energy
function for the same ensembles (the AGBNP1 ensemble and the explicit
solvent ensemble) showed marked improvements (see
Fig.~\ref{agbnp1de.fig}, second and third row). The second row of
graphs in Fig.~\ref{agbnp1de.fig} (panels D--F) correspond to the
AGBNP2 model without hydrogen bonding to solvent corrections (referred
to as AGBNP2-SEV). We see that the introduction of the SEV model for
the solute volume already drastically lowers the energy scores of the
explicit solvent conformations relative to the implicit solvent
conformations. The AGBNP2-SEV energy gaps between the explicit and
implicit solvation ensembles are reduced approximately by a factor of
two for all three mini-proteins, moreover the energy distributions for
the two ensembles now partially overlap.  The AGBNP2/OPLS-AA energy
scores with the hydrogen bonding corrections and the parameters in
Table \ref{tab.hs} (Figs.~\ref{agbnp1de.fig}G--I) favor the explicit
solvent ensemble relative to the AGBNP1 ensemble even further. The
average energy gaps between the two ensembles are reduced to 1.3, 1.0,
and 1.2 kcal/mol per residue for, respectively, the trp-cage, cdp-1,
and fsd-1 mini-proteins, compared to the average energy gaps of 3.3,
4.4, and 5.7 kcal/mol with the AGBNP1/OPLS-AA effective potential (a 3
to 5-fold improvement) and the overlaps between the energy
distributions of the implicit and explicit solvation ensembles is
further increased relative to AGBNP2-SEV
(compare Figs.~\ref{agbnp1de.fig}G--I with
Figs.~\ref{agbnp1de.fig}D--F). The energy bias per residue of AGBNP2
relative to the explicit solvent ensemble is comparable to thermal
energy and smaller than the spread of the energy distributions
indicating that AGBNP2 is capable of generating  explicit solvent-like
conformations much more frequently than AGBNP1.

\begin{figure}[tbp]
\begin{center}
\scalebox{0.75}{\includegraphics{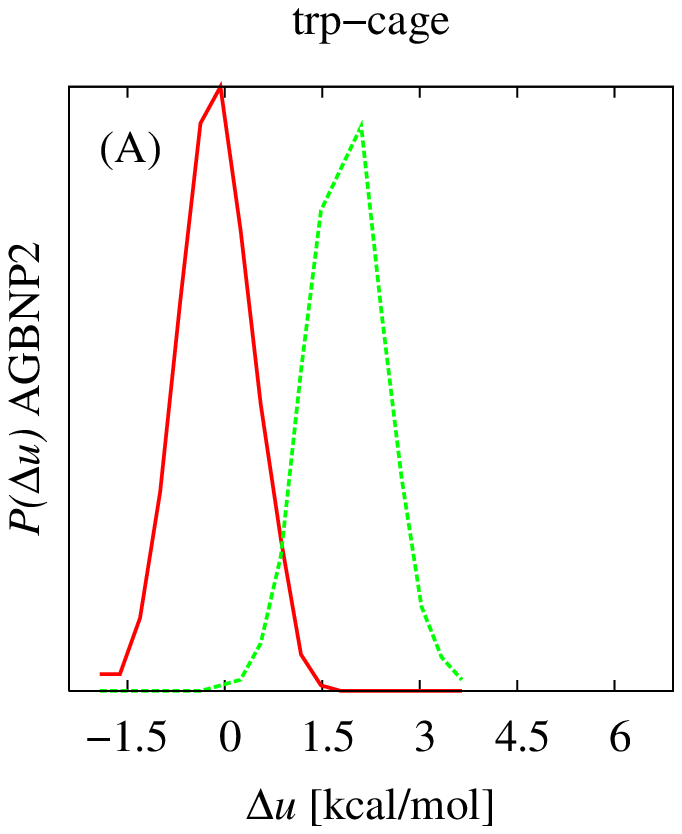}\hspace{-2.45in}\includegraphics{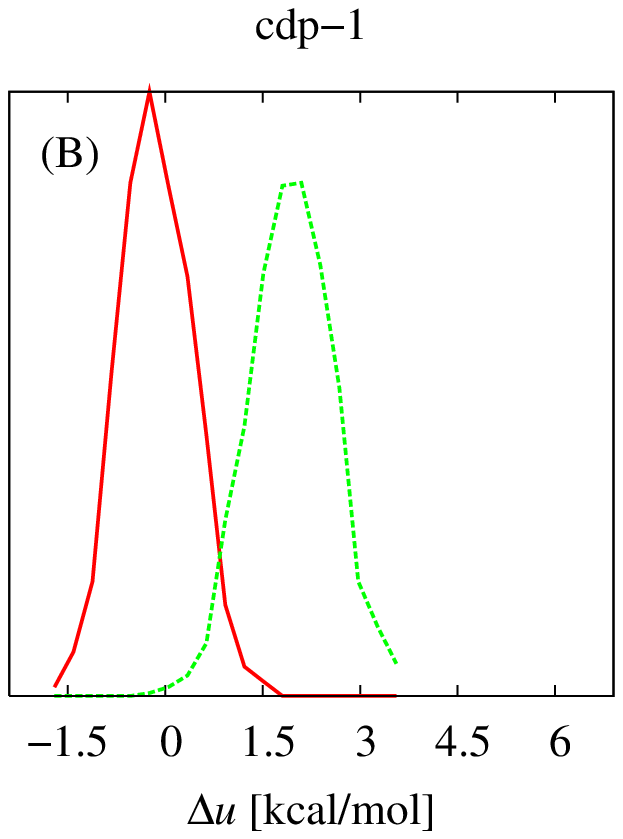}\hspace{-2.5in}\includegraphics{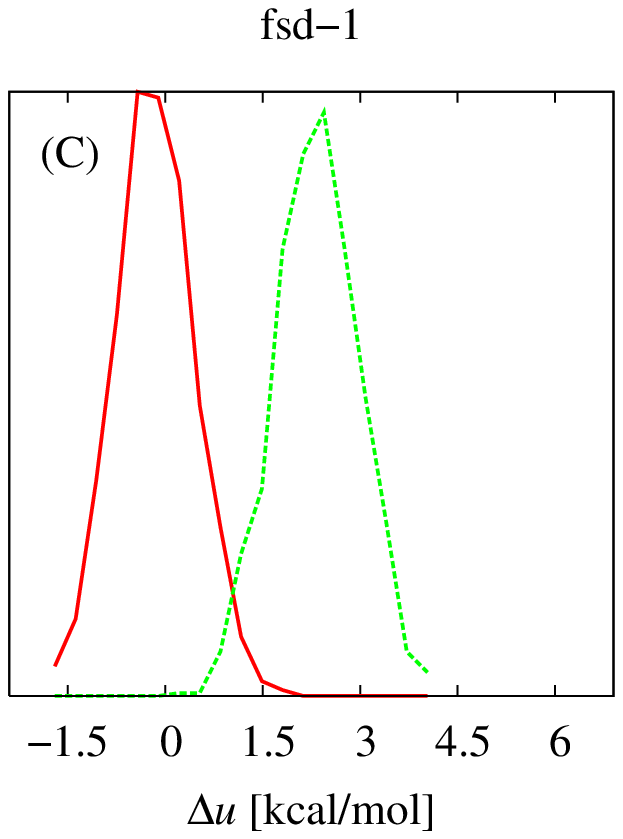}}
\end{center}
\caption{\label{agbnp2de.fig} {\small Probability distributions of the
AGBNP2/OPLS-AA effective potential energy for the conformational
ensembles of the trp-cage (A), cdp-1 (B), and fsd-1 (C) mini-proteins
obtained using the present AGBNP2 solvation model (full line) and
explicit solvation (dashed line). The distributions are shown versus
the energy gap per residue ($\Delta u$) relative to the mean effective
potential energy of the implicit solvent ensemble.  } }
\end{figure}

The energy scoring experiments on the explicit solvent and AGBNP1
ensembles described above were very useful for tuning the formulation
of the AGBNP2 model without requiring running lengthy MD
simulations. They do not, however, guarantee that the conformational
ensembles generated with the AGBNP2 solvation model will more closely
match the explicit solvent ensembles than those generated with
AGBNP1. This is because the new solvation model could introduce new
energy minima not encountered with AGBNP1 or explicit solvation that
would be visited only by performing conformational sampling with
AGBNP2. To validate the model in this respect we obtained MD
trajectories with the AGBNP2 implicit solvent model and compared the
corresponding probability distributions of the effective energy with
those of the explicit solvent ensembles similarly as above.  The
results for the three mini-proteins, shown in Fig.~\ref{agbnp2de.fig},
indicate that the AGBNP2-generated ensembles display significantly
smaller bias (mean energy gaps per residue of 2.0, 2.1, and 2.5
kcal/mol for, respectively, the trp-cage, cdp-1, and fsd-1
mini-proteins) than AGBNP1 (Fig.~\ref{agbnp1de.fig}, panels A--C)
which instead yielded energy gaps of $3.3$, $4.4$, and $5.7$ kcal/mol
per residue respectively. This observation confirms that
conformational sampling with AGBNP2 produces conformations that more
closely match on average the explicit solvent ensemble without
producing unphysical minima that deviate significantly from it.

We have structurally analyzed the conformational ensembles obtained
with the AGBNP1 and AGBNP2 models to establish the degree of
improvement achieved with the new model in terms of occurrence of
intramolecular interactions. The salient results of this analysis are
shown in Table \ref{hbt.table}. This table reports for each
mini-protein the average number of intramolecular hydrogen bonds and
ion pairs. The number of hydrogen bonds is further decomposed into
those involving only polar groups (including the backbone) and those
involving a polar groups and the sidechain of a charged residue
(arginine, lysine, aspartate, and glutamate). As noted above it is
apparent from this data that the AGBNP1 model produces conformations
with too many hydrogen bonds and ion pairs.  The majority of the
excess hydrogen bonds with AGBNP1 involve aminoacid sidechains.
Similarly, too many ion pairs are observed in the AGBNP1 ensemble
particularly for the fsd-1 mini-protein ($4.6$ ion pairs on average
with AGBNP1 compared to only $1.4$ in explicit solvent). The AGBNP2
ensembles, in comparison, yields considerably fewer intramolecular
hydrogen bonds. For instance, the average number of hydrogen bonds for
cdp-1 is reduced from $24.5$ with AGBNP1 to $15.4$ with AGBNP2. With
AGBNP2 the number of polar--polar hydrogen bonds is generally in good
agreement with explicit solvation. However the greatest improvement is
observed with polar--charged interactions. For example the number of
polar--charged hydrogen bonds of fsd-1 is reduced by almost ten fold
in going from AGBNP1 to AGBNP2 to reach good agreement with explicit
solvation. Importantly, a significant fraction of the excess
polar--charged interactions observed with AGBNP1 corrected by AGBNP2
are interactions between the peptide backbone and charged sidechains
that would otherwise interfere with the formation of secondary
structures.

With AGBNP2 we observe small but promising improvements in terms of
ion pair formation. The average number of ion pairs of cdp-1
consistently agrees between all three solvation models, and the only
possible ion pair in trp-cage is more stable in both implicit solvent
formulations than in explicit solvent (it occurs in virtually all
implicit solvent conformations compared to only 30\% of the
conformations in explicit solvent). However the average number of ion
pairs for fsd-1 is reduced from $4.6$ with AGBNP1 to $4.0$ with
AGBNP2. We observe good agreement between the number of ion pairs
involving lysine with either AGBNP1 or AGBNP2 and explicit
solvation. However ion pairs involving arginine are generally more
stable with implicit solvation than with explicit solvation. The
agreement in the number of ion pairs with cdp-1 is due to the fact
that for this mini-protein the two possible ion pairs involving
arginine result stable with explicit solvation as well as with
implicit solvation. For the other two
mini-proteins however, ion pairs involving arginine that are
marginally stable with explicit solvation are found to be
significantly more stable with implicit solvation, although less so
with AGBNP2 solvation.

\section{Discussion}

The non-linear and asymmetric dielectric response of water stems
primarily by the finite extent and internal structure of water
molecules.\cite{Levy:Gallicchio:98} As further discussed below, the
modeling of effects due to water granularity is important for the
proper description of molecular association equilibria.  Integral
equation methods\cite{Yoshida:Imai:Kovalenko:Hirata:2009} provide an
accurate implicit solvation description from first principles, however,
despite recent progress,\cite{Miyata:Hirata:2008} they are not yet
applicable to molecular dynamics of biomolecules. The primary aim of
the present study has been to formulate an analytical and
computational efficient implicit solvent model incorporating solvation
effects beyond those inherent in standard continuum dielectric models
and, by so doing, achieving an improved description of solute
conformational equilibria.

In this work we have developed the AGBNP2 implicit solvent model
which is based on an empirical (but physically-motivated) first
solvation shell correction function parametrized against experimental
hydration free energies of small molecules and the results of explicit
solvent molecular dynamics simulations of a series of
mini-proteins. The correction function favors conformations of the
solute in which polar groups are oriented so as to form hydrogen bonds
with the surrounding water solvent thereby achieving a more balanced
equilibrium with respect to the competing intramolecular hydrogen bond
interactions. A key ingredient of the model is an analytical
prescription to identify and measure the volume of hydration sites on
the solute surface. Hydration sites that are deemed too small to
contain a water molecule do not contribute to the solute hydration
free energy. Conversely, hydration sites of sufficient size form
favorable interactions with nearby polar groups. This model thus
incorporates the effects of both water granularity and of non-linear
first shell hydration effects. 

The GB and non-polar models in the AGBNP2 implicit solvent model
provide the linear continuum dielectric model basis of the model as
well as a description of non-electrostatic hydration
effects.\cite{Gallicchio:Levy:2004} In this work the GB and
solute--solvent dispersion interaction energy models are further
enhanced by replacing the original van der Waals solute volume model
with a more realistic solvent excluding volume (SEV) model. The new
volume description improves the quality of the Born radii of buried
atoms and atoms participating in intramolecular interactions which
would otherwise be underestimated due to high dielectric interstitial
spaces present with the van der Waals volume
description.\cite{Swanson:Mongan:McCammon:2005} GB models with these
characteristics have been previously proposed.  The GBMV series of
models\cite{Lee:Salsbury:Brooks:2002,Lee:Feig:Salsbury:Brooks:2003,Chocholousova:Feig:2006}
represent the SEV on a grid which, although accurate, is
computationally costly and lacks frame of reference invariance.  The
pairwise descreening-based GB$^{\mathrm OBC}$
model\cite{Onufriev:Bashford:Case:2004} introduced an empirical
rectifying function to increase the Born radii of buried atoms in an
averaged, geometry-independent manner. The GBn
model\cite{Mongan:Simmerling:McCammon:Case:Onufriev:2007} introduced
the neck region between pairs of atoms as additional source of
descreening, dampened by empirical parameters to account in an average
way for overlaps between neck regions and between solute atoms and
neck regions.  The approach proposed here to represent the SEV
consists of computing the atomic self-volumes [Eq.\ (\ref{Vself})],
used in the pairwise descreening computation, using enlarged atomic
radii so as to cover the interatomic interstitial spaces. The
self-volume of each atom is then reduced proportionally to its solvent
accessible surface area [see Eq.\ (\ref{sj_def})] to subtract the
volume in van der Waals contact with the solvent. We show (Fig.\
\ref{br.fig}) that this model reproduces well Born radii computed from
an accurate numerical representation of the SEV, noting that
improvements for the Born radii of atoms in loosely packed hydrophobic
interior, while significant, are still not optimal.  Although
approximate, this representation of the SEV maintains the simplicity
and computational efficiency of pairwise descreening schemes, while
accounting for atomic overlaps in a consistent and parameter-free
manner.

The new AGBNP2 model has been formulated to be employed in molecular
dynamics conformational sampling applications, which require potential
models of low computational complexity and favorable scaling
characteristics, and with analytical gradients.  These requirements
have posed stringent constraints on the design of the model and the
choice of the implementation algorithms. In the formulation of AGBNP2
we have reused as much as possible well-established and efficient
algorithms developed earlier for the AGBNP1 model. For example the key
ingredient of the hydrogen bonding correction function is the free
volume of an hydration site, which is computed using a methodology
developed for AGBNP1 to compute atomic self-volumes. Similarly the
SEV-based pairwise descreening procedure employs atomic surface areas
(see Eq.\ \ref{sj_def}), computed as previously
described.\cite{Gallicchio:Levy:2004} AGBNP2 suffers additional
computational cost associated with the SEV-based pairwise descreening
procedure and the hydrogen bonding correction function. This handicap
however is offset by having only one solute volume model in AGBNP2
rather than two in AGBNP2. AGBNP1 requires two separate invocations of
the volume overlaps machinery [Eq.\ \ref{V1}] for each of the two
volume models it employs, corresponding to the van der Waals atomic
radii for the pairwise descreening calculation and enlarged radii for
the surface area calculation.\cite{Gallicchio:Levy:2004} AGBNP2
instead employs a single volume model for both the pairwise
descreening and surface area calculations.  A direct CPU timing
comparison between the two models can not be reported at this time
because the preliminary implementation of the AGBNP2 computer code
used in this work lacks key data caching optimizations similar to
those already employed in AGBNP1. Given the computational advantages
of the new model discussed above we expect to eventually obtain
similar or better performance than with AGBNP1.

The AGBNP2 model has been parametrized against experimental hydration
free energies of a series of small molecules and with respect to the
ability of reproducing energetic and structural signatures of the
conformational ensemble of three mini-proteins generated with explicit
solvation. These data sources are chosen so as to ensure that the
resulting model would be applicable to both hydration free energy
estimation and conformation equilibria, which are fundamental
characteristics for models aimed at protein-ligand binding affinity
estimation.  On the other hand, experimental hydration
free energies and explicit solvent conformational ensembles are to
some extent incompatible with one another given the limitations of
even the best fixed charge force fields and explicit solvation models
to reproduce experimental hydration free energies of small molecules
with high accuracy.\cite{Deng:Roux:2004,Shirts:Pande:2005,Mobley:Bayly:Dill:2009} Mindful of this issue we did not seek a
perfect correspondence with the experimental hydration free energy
values. We first obtained parameters by fitting against the small
molecule experimental hydration free energies, then adjusted the
parameters to improve the agreement with the explicit solvent data
making sure that the predicted small molecule hydration free energies
remained within a reasonable range relative to the experimental
values. In practice this procedure yielded predicted hydration free
energies in good agreement with the experimental values with the
exception of the carboxylate and guanidinium ions (see Table
\ref{tab.small.mol}) for which AGBNP2 predicts more favorable
hydration free energies than the experiments; a consequence of the
large hydrogen bonding corrections necessary to reduce the occurrence
of intramolecular electrostatic interactions in the investigated
proteins. As discussed further below, limitations in the description
of hydration sites adopted for carboxylate and guanidinium ions may be
partly the cause of the observed inconsistencies for these functional
groups.

Parametrization and validation of the model focused in part on
comparing the effective potential energy distributions of implicit
solvent conformational ensembles with those of explicit solvent
ensembles. We observed that the AGBNP1 solvation model energetically
ranked explicit solvent conformations significantly less favorably
than implicit solvent conformations.  The AGBNP2 model is
characterized by smaller energetic bias relative to the explicit
solvent ensembles, indicating that conformational sampling with the
AGBNP2/OPLS-AA energy function produces conformations that more
loosely match those obtained with explicit solvation.  This result is
a direct consequence of employing the more realistic solvent excluding
volume description of the solute which yields larger Born radii for
buried groups, as well as the hydrogen bonding to solvent corrections,
which favor solvent exposed conformations of polar groups.
Furthermore, comparison of the energy distributions of the AGBNP2 and
explicit solvent ensembles for the three mini-proteins (Fig.\
\ref{agbnp2de.fig}) shows, in contrast to the AGBNP1 results, that the
AGBNP2 bias for the two more charge-rich mini-proteins (cdp-1 and
fsd-1) is similar to that of the least charged one (trp-cage). This
suggests that the residual energetic bias of the AGBNP2 model is
probably related to the non-polar model rather than the electrostatic
model. Future studies will address this particular aspect of the
model.

The energy scoring studies conducted in this work indicate that AGBNP2
is a significant improvement over AGBNP1. They also show, however,
that the new model falls short of consistently scoring explicit
solvent conformations similarly to implicit solvent conformations.
Although an optimal match between energy distributions is a necessary
condition for complete agreement between implicit and explicit
solvation results, it is unrealistic to expect to reach this ultimate
goal at the present level of model simplification.  Increasing the
magnitude of the hydrogen bonding corrections can improve the
agreement between the explicit and implicit solvation energy
distributions, albeit at the expense of the quality of the predicted
small molecule hydration free energies.  It seems likely that the no
parametrization of the current model would yield both good relative
conformational free energies and hydration free energies.  Future work
will pursue the modeling of additional physical and geometrical
features necessary to improve the agreement between implicit and
explicit solvation energy distributions.  The energy gap between the
implicit solvent and explicit solvent energy distributions used here
is, we believe, a meaningful measure of model quality and could serve
as a useful general validation tool to compare the accuracy of
implicit solvent models.

The excessive number of intramolecular electrostatic interactions
involving charged groups has been one of the most noticeable
shortcoming of GB-based implicit solvent models.\cite{Zhou:Berne:2002}
To correct this tendency we have in the past adopted in the AGBNP1
model ad-hoc strategies aimed at either destabilizing electrostatic
intramolecular interactions\cite{Felts:Harano:Gallicchio:Levy:2004}
or, alternatively, stabilizing the competing solvent-separated
conformations.\cite{Felts:Gallicchio:Levy:2008} This work follows the
latter approach using a more robust and physically-motivated framework
based on locating and scoring hydration sites on the solute surface as
well as adopting a more realistic volume model. Structural
characterization of the conformational ensembles has shown that AGBNP2
produces significantly fewer intramolecular interactions than AGBNP1
reaching good agreement with explicit solvent results. The reduction
of intramolecular interactions has been the greatest for interactions
between polar and charged groups. We believe that the excessive
tendency toward the formation of intramolecular interactions to be the
root cause of the high melting temperatures of structured
peptides\cite{Weinstock:Narayanan:Baum:Levy:2007} predicted with
AGBNP1. Given the reduction of intramolecular interactions achieved
with AGBNP2, we expect the new model to yield more reasonable peptide
melting temperatures; a result which we hope to report in future
publications.

Less visible improvements have been obtained for ion pairs involving
arginine sidechains which remain more stable with implicit solvation
than explicit solvation. However, significantly, with AGBNP2 we
observed a more dynamical equilibrium between ion-paired and
solvent-separated conformations of arginine sidechains that was not
observed with AGBNP1.  This result is promising because it indicates
that the AGBNP2 solvation model, although still favoring ion paired
conformations, produces a more balanced equilibrium, which is instead
almost completely shifted towards contact conformations with AGBNP1.
Nevertheless it is apparent that the AGBNP treatment of the
guanidinium group of arginine is not as good as for other groups. This
limitation appears to be shared with other functional groups
containing sp2-hybridized nitrogen atoms as evidenced for example by
the relatively lower quality of the hydration free energy predictions
for amides and nitrogen-containing aromatic compounds (Table
\ref{tab.small.mol}).  Similar implicit solvent overstabilization
solvation of arginine-containing ion pairs has been observed Yu et
al.\cite{Yu:Jacobson:Friesner:2004} in their comparison of Surface
Generalized Born (SGB) and SPC explicit solvation with the OPLS-AA
force field. Despite quantitative differences, the explicit solvent
studies (with the TIP3P water model) of Masunov and
Lazaridis\cite{Masunov:Lazaridis:2003} and Hassan\cite{Hassan:2004},
using the CHARMM force field, and that of Mandell at
al.\cite{Mandell:Jacobson:2007}, using the OPLS-AA force field, have
confirmed that arginine forms the strongest ion pairing interactions,
especially in the bidentate coplanar conformation.  These observations
are consistent with the present explicit solvent results using OPLS-AA
and the SPC water model, where we find that most of the ion pairs of
the mini-proteins were found to involve arginine sidechains. In
contrast to our present implicit solvent results, however, the work of
Masunov and Lazaridis\cite{Masunov:Lazaridis:2003} indicated that the
GB-based implicit solvent model that they
analyzed\cite{Dominy:Brooks:99} produced potentials of mean force
for arginine-containing ion pairs in general agreement with explicit
solvation.

\begin{figure}[tbp]
\begin{center}
\hspace{-0.7in}\scalebox{0.9}{\includegraphics{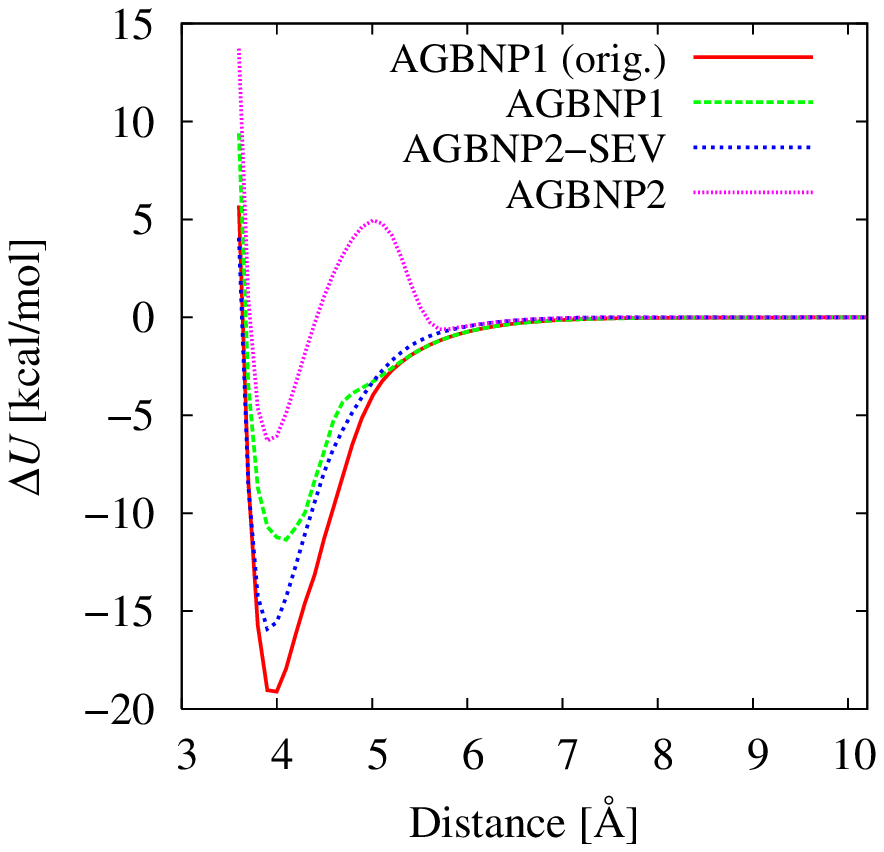}\hspace{-1.45in}\includegraphics{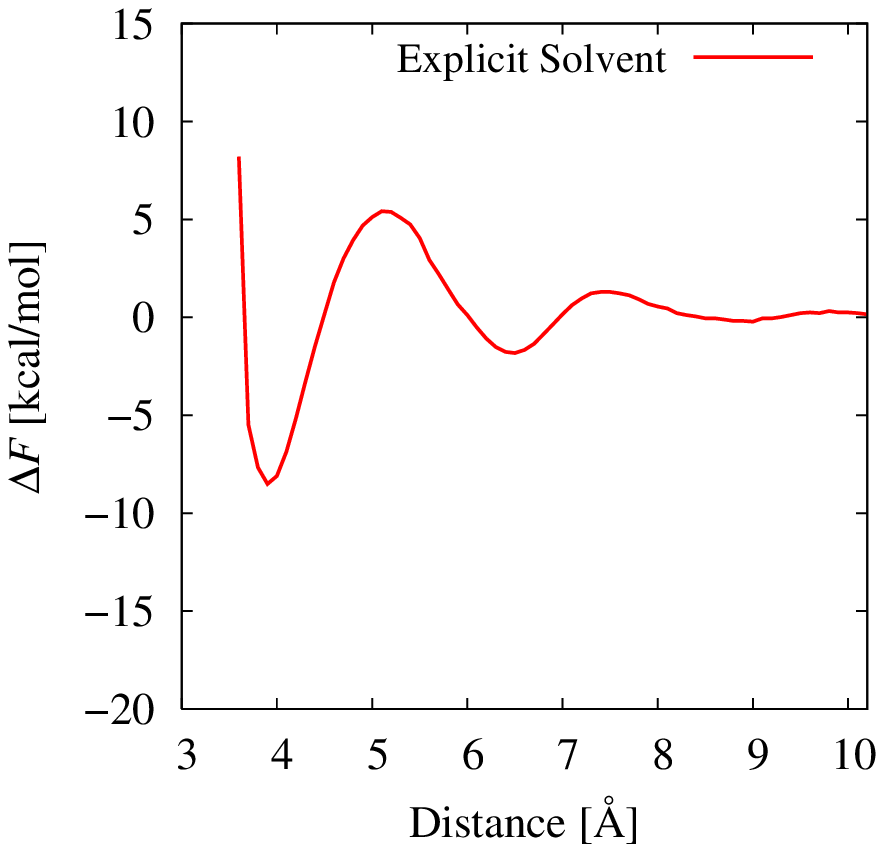}}

\caption{\label{pmfs1.fig} {\small Potential of mean force of ion
    pair formation between propyl-guanidinium and ethyl-acetate in the
    coplanar orientation with AGBNP implicit solvation (A) and
    explicit solvation (B; reference
    \cite{Mandell:Jacobson:2007}). In (A) ``AGBNP1 (orig.)'' refers to the
    original AGBNP1 parametrization\cite{Gallicchio:Levy:2004},
    ``AGBNP1'' refers to the AGBNP1 model used in this work which includes a surface tension parameter correction for the
    carboxylate group aimed at reducing the occurrence of ion
    pairs,\cite{Felts:Gallicchio:Levy:2008} 
    ``AGBNP2'' refers to the current model and ``AGBNP2-SEV'' the
    current model without hydrogen bonding corrections. The potentials
    of mean force are plotted with respect to the distance between the
    atoms of the protein sidechain analogs equivalent to the C$\zeta$ of arginine and the C$\gamma$ of aspartate.
}}
\end{center}
\end{figure}

To rationalize the present implicit solvent results concerning ion
pair formation, it has been instructive to analyze the potentials of
mean force (PMF) of ion pair association with the AGBNP model. As an
example, Fig.\ \ref{pmfs1.fig} shows 
the PMF for the association of propyl-guanidinium (arginine sidechain
analog) and ethyl-acetate (aspartate and glutamate analog) in a
bidentate coplanar conformation (similar to the arrangement used
previously\cite{Masunov:Lazaridis:2003,Hassan:2004,Yu:Jacobson:Friesner:2004,Mandell:Jacobson:2007}) for various AGBNP implementations.
The corresponding explicit solvent PMF obtained by
Mandell et al.\cite{Mandell:Jacobson:2007} is also shown in Fig.\
\ref{pmfs1.fig} for comparison.  The original AGBNP1
parametrization\cite{Gallicchio:Levy:2004} clearly leads to an overly
stable salt bridge with the contact conformation scored at
approximately $-19$ kcal/mol relative to the separated conformation,
compared with $-8.5$ kcal/mol with explicit solvation. The AGBNP1
parametrization analyzed here, which includes an empirical surface
area correction to reduce the occurrence of ion
pairs,\cite{Felts:Gallicchio:Levy:2008} yields a contact free energy ($-11$ kcal/mol)
in much better agreement with explicit solvation, although the shape
of the PMF is not properly reproduced. The present AGBNP2 model
without hydrogen bonding corrections (labeled ``AGBNP2-SEV'' in Fig.\
\ref{pmfs1.fig}) yields a PMF intermediate between the original and
corrected AGBNP1 parametrization. The AGBNP2 model with hydrogen
bonding corrections yields the PMF with the closest similarity with
the one obtained in explicit solvent (Fig.\ \ref{pmfs1.fig}B). Not only the contact free energy ($-6.5$
kcal/mol) is in good agreement with the explicit solvent result,
but, importantly, it also reproduces the solvation barrier of the PMF at $5$
\AA\ separation, corresponding to the distance below which there is
insufficient space for a water layer between the ions.

It is in this range of distances that the greatest discrepancies are
observed between PMF's with explicit solvation and some GB-based implicit
solvation models\cite{Masunov:Lazaridis:2003,Yu:Jacobson:Friesner:2004} that
do not model effects due to the finite size of water molecules.
Both the hydrogen bonding correction and the SEV volume
description employed in AGBNP2, which are designed to take into account the
granularity of the water solvent -- the hydrogen bonding correction
through the minimum free volume of water sites [Eq.\
(\ref{HB_corr_energy})] and the SEV model through the water radius
offset [Eq.\ (\ref{big_radii})] -- make it possible to
reproduce the solvation barrier typical of molecular association
processes in water. 

The good correspondence between the AGBNP2 and explicit solvent PMF's
for propyl-guanidinium and ethyl-acetate (Fig.\ \ref{pmfs1.fig}) stand
in contrast with the residual AGBNP2 overprediction of arginine salt
bridges compared to explicit solvation (Table \ref{hbt.table}). We
observed that in the majority of arginine salt bridges occurring with
AGBNP2, the guanidinium and carboxylate groups interact at an angle
rather than in the coplanar configuration discussed above. We have
confirmed that the PMF of ion pair formation for an angled
conformation (not shown) indeed shows a significantly more attractive
contact free energy than the coplanar one.  This result indicates that
the in-plane placement of the hydration sites for the carboxylate
groups (see Appendix) does not sufficiently penalize angled ion pair
arrangements. This observation is consistent with the need of
introducing of an isotropic surface-area based hydration correction
for carboxylate groups (the reduced $\gamma$ parameter for the
carboxylate oxygen atoms), which showed some advantage in terms of
reducing the occurrence of salt bridges. Future work will focus on
developing a more general hydration shell description for carbonyl
groups and related planar polar groups to address this issue.

\section{Conclusions}

We have presented the AGBNP2 implicit solvent model, an evolution of
the AGBNP1 model we have previously reported, with the aim of
incorporating modeling of hydration effects beyond the continuum
dielectric representation. To this end a new hydration free energy
component based on a
procedure to locate and score hydration sites on the solute surface is
used to model first solvation shell effects, such as hydrogen bonding,
which are poorly described by continuum dielectric models. This new
component is added to the Generalized Born and non-polar AGBNP models
which have been improved with respect to the description of the solute
volume description. We have introduced an analytical Solvent Excluding
Volume (SEV) model which reduces the effect of artefactual
high-dielectric interstitial spaces present in conventional van der
Waals representations of the solute volume. The new model is
parametrized and tested with respect to experimental hydration free
energies and the results of explicit solvent simulations.
The modeling of the
granularity of water is one of the main principles employed in the
design of the empirical first shell solvation function and the SEV
model, by requiring that hydration sites have a minimum
available volume based on the size of a water molecule. We show that
the new volumetric model produces Born radii and surface areas in good
agreement with accurate numerical evaluations. The results of
Molecular Dynamics simulations of a series of mini-proteins show that
the new model produces conformational ensembles in much better
agreement with reference explicit solvent ensembles than the AGBNP1
model both with respect to structural and energetics measures.

Future development work will focus on improving the modeling of
some functional groups, particularly ionic groups involving sp2
nitrogen, which we think are at the basis of the residual
excess occurrence of salt bridges, and on the optimization of the
AGBNP2 computer code implementation. We also plan to test the model on
wide variety of benchmarks including peptide folding.

\appendix

\section{Hydration Site Locations}

\begin{figure}[tbp]
\begin{center}
\scalebox{1.0}{\includegraphics{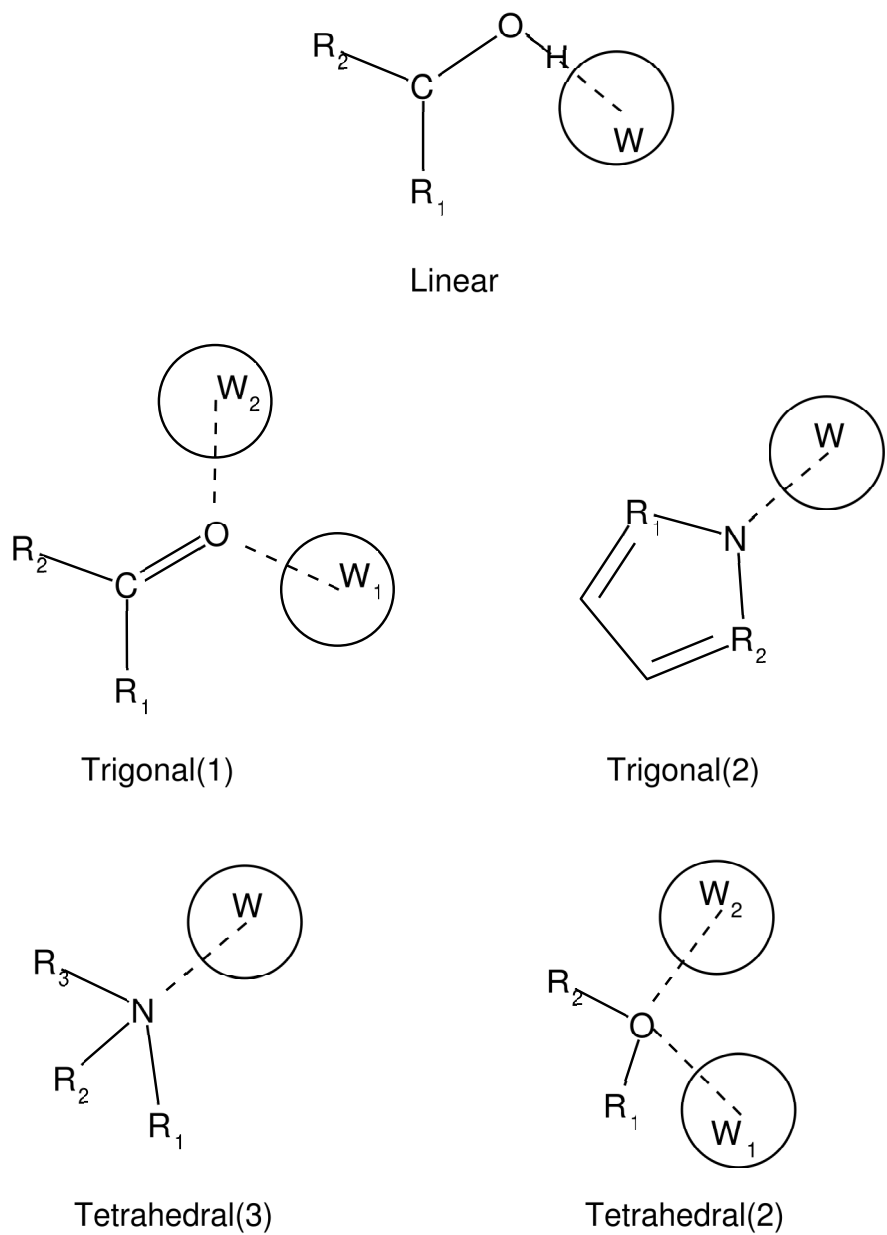}}
\end{center}
\caption{\label{water_sites.fig}
Diagram illustrating the hydration site locations for each of the functional
group geometries used in this work. Linear: hydrogen bond donor;
trigonal(1) and trigonal(2): trigonal planar geometries with,
respectively, one and two covalent bonds on the acceptor atom;
tetrahedral(2) and tetrahedral(3): tetrahedral geometries with,
respectively, two and three covalent bonds on the acceptor
atom. Representative molecular structures are shown for each geometry.
}
\end{figure}

Fig.~\ref{water_sites.fig} shows the location of the hydration sites
for the functional groups listed in Table~\ref{tab.hs}. Each hydration site
is represented by a sphere of $1.4$ \AA\ radius. The distance
$d_{\rm HB}$ between the donor or acceptor heavy atom and the center
of the hydration site sphere is set to $2.5$ \AA.

There is a single linear geometry for HB donor groups. The
corresponding hydration site is placed at a distance $d_{\rm HB}$ from
the heavy atom donor along the heavy atom-hydrogen bond.

Acceptor trigonal geometries have one or two hydration sites depending
on whether the acceptor atom is bonded to, respectively, two or one other atom. In the former case the water site is placed along
the direction given by the sum of the unit vectors corresponding to
the sum of the NR$_1$ and NR$_2$ bonds (following the atom names in
Fig.~\ref{water_sites.fig}). In the latter case the W$_1$ site (see
Fig.~\ref{water_sites.fig}) is placed in the R$_1$CO plane forming an
angle of $120^\circ$ with the CO bond. The W$_2$ site is placed
similarly.

Acceptor tetrahedral geometries have one or two hydration sites depending
on whether the acceptor atom is bonded, respectively, to three or two
other atoms. In the former case the water site is placed along
the direction given by the sum of the unit vectors corresponding to
the sum NR$_1$, NR$_2$, and NR$_3$ bonds. In the latter case the
positions of the W$_1$ and W$_2$ sites are given by
\begin{eqnarray}
{\mathbf w}_1 & = & \mathbf{O} + d_{\rm HB} \left( \cos\theta \mathbf{u}_1
+ \sin\theta  \mathbf{u}_2 \right) \nonumber \\ 
{\mathbf w}_1 & = & \mathbf{O} + d_{\rm HB} \left( \cos\theta \mathbf{u}_1
- \sin\theta  \mathbf{u}_2 \right) \nonumber
\end{eqnarray}
where $\mathbf{O}$ is the position of the acceptor atoms, $\theta =
104.4^\circ$, and $\mathbf{u}_1$ and $\mathbf{u}_2$ are, respectively,
the unit vectors corresponding to the OR$_1$ and OR$_2$ bonds.

\section{Gradients of GB and van der Waals Energies}

The component of the gradient of the AGBNP2 van der Waals energy at constant
self-volumes is the same as in the AGBNP1 model [see Appendix C of
  reference \cite{Gallicchio:Levy:2004}]. In AGBNP2 the expression for the
component of the gradient corresponding to variations in the atomic scaling
factors, $s_{ij}$, includes pair corrections at all overlap levels because of
the presence of multi-body volumes in $V''_{ij}$. In addition, a new
component corresponding to the change in surface areas appears: 
\begin{eqnarray}
\left( \frac{\partial \beta_j}{\partial \mathbf{r}_i}\right)_Q & = & 
- \frac{1}{4 \pi} \sum_k \frac{\partial s_{kj}}{\partial \mathbf{r}_i}
Q_{kj} \nonumber \\
& = & - \frac{1}{4 \pi} \sum_k \frac{1}{V_k} 
\frac{\partial V'_k}{\partial \mathbf{r}_i} Q_{kj} \label{dbeta_dr1} \\
& &   - \frac{1}{4 \pi} \sum_k \frac{1}{V_k} 
\frac{\partial  V'_{kj}}{\partial \mathbf{r}_i} Q_{kj} \label{dbeta_dr2} \\
& &   + \frac{1}{4 \pi} \sum_k \frac{1}{V_k} p_k 
\frac{\partial A_k}{\partial \mathbf{r}_i} Q_{kj} \, . \label{dbeta_dr3}
\end{eqnarray}
Eq.~(\ref{dbeta_dr1}) leads to the same expression of the derivative
component as in the AGBNP1 model [Eq.~(72) in reference
  \cite{Gallicchio:Levy:2004}] (except for
the extra elements in the 2-body terms due to the inclusion of the
$1/2 V_{kj}$ correction term). Eq.~(\ref{dbeta_dr2}) corresponds to the
component of the derivative due to variations in $V'_{jk}$, the
volume to be added to the self-volumes of $j$ and $k$ to obtain the
$s_{jk}$ and $s_{kj}$ scaling factors. In the AGBNP1 model this
component included only 2-body overlap volumes; in AGBNP2 this term
instead includes overlap volumes
at all orders. Finally, Eq.~(\ref{dbeta_dr3}), where $A_k$ is the
surface area of atom $k$, leads to the component of the derivatives of
the GB and vdW terms due to variations of the exposed surface
area. The latter two terms are new for AGBNP2.

\subsection{Component of derivative from Eq.~(\ref{dbeta_dr2})}

From Eq.~(63) in reference \cite{Gallicchio:Levy:2004} and Eq.~(\ref{dbeta_dr2}) we have
\begin{equation}
-4 \pi \left( \frac{\partial \Delta G_{vdW}}{\partial \mathbf{r}_i}\right)_{Q2} = \sum_{jk} W_{kj} 
\frac{\partial V'_{kj}}{\partial \mathbf{r}_i}\, , \label{dGvdw_Q2}
\end{equation}
where $W_{kj}$ has the same expression as in Eq.~(69) in reference
\cite{Gallicchio:Levy:2004}.  In working with Eq.~(\ref{dGvdw_Q2}) it is important to note that,
whereas $V'_{kj}$ is symmetric with respect to swapping the $j$ and
$k$ indexes, $W_{kj}$ and $W_{jk}$ are different from each
other. Substituting Eq.~(\ref{Vpp_def}) into Eq.~(\ref{dGvdw_Q2}) and
expanding over symmetric terms we obtain
\begin{eqnarray}
- 4 \pi \left( \frac{\partial \Delta G_{vdW}}{\partial
  \mathbf{r}_i}\right)_{Q2} & = &
 \frac12 \sum_{jk} W_{kj} \frac{\partial V_{kj}}{\partial \mathbf{r}_i}
  \nonumber \\
&& -\frac13 \sum_{jkl} W_{kj} \frac{\partial V_{jkl}}{\partial
  \mathbf{r}_i}
\nonumber \\
&& + \frac12 \frac14 \sum_{jklp} W_{kj} \frac{\partial V_{jklp}}{\partial
  \mathbf{r}_i} 
\nonumber \\
&& - \ldots \label{dGvdW_ri1}
\end{eqnarray}
Eq.~(\ref{dGvdW_ri1}) is simplified by noting that
\begin{equation}
\frac{\partial V_{jk\ldots}}{\partial \mathbf{r}_i} =
\delta_{ij}  \frac{\partial V_{ik\ldots}}{\partial \mathbf{r}_i} +
\delta_{ik}  \frac{\partial V_{ji\ldots}}{\partial \mathbf{r}_i} +
\ldots \label{p_deltas}
\end{equation}
Eq.~(\ref{p_deltas}) is inserted in Eq.~(\ref{dGvdW_ri1}) and sums are
reduced accordingly, then symmetric
terms are collected into single sums by re-indexing the
summations, obtaining
\begin{eqnarray}
- 4 \pi \left( \frac{\partial \Delta G_{vdW}}{\partial
  \mathbf{r}_i}\right)_{Q2} & = &
 \frac12 \sum_j (W_{ij} + W_{ji}) 
\frac{\partial V_{ij}}{\partial \mathbf{r}_i} \nonumber \\
&&- \frac13 \sum_{j<k} \left[ (W_{ij} + W_{ji}) + (W_{jk} + W_{kj}) + 
(W_{ik} + W_{ki}) \right]
\frac{\partial V_{ijk}}{\partial \mathbf{r}_i} \nonumber \\
&&+ \frac14 \sum_{j<k<l} \left[
(W_{ij} + W_{ji}) + (W_{ik} + W_{ki}) + 
(W_{il} + W_{li}) + \right. \nonumber \\
&& \left.
(W_{jk} + W_{kj}) + (W_{jl} + W_{lj}) + 
(W_{kl} + W_{lk})
\right]
\frac{\partial V_{ijkl}}{\partial \mathbf{r}_i} \nonumber \\
&&- \ldots \label{dGvdW_ri2}
\end{eqnarray} 
The corresponding expression for the gradient of $\Delta G_{\rm GB}$
is similar but employs the $U_{ij}$ factors of Eq.~(78) of reference
\cite{Gallicchio:Levy:2004} rather than $W_{ij}$.

\subsection{Component of derivative from Eq.~(\ref{dbeta_dr3})}

Inserting Eq.~(\ref{dbeta_dr3}) in Eq.~(63) of reference
\cite{Gallicchio:Levy:2004} gives
\[
4 \pi \left( \frac{\partial \Delta G_{\rm vdW}}{\partial \mathbf{r}_i}
\right)_{Q3} =
\sum_{jk} W_{kj} p_k \frac{\partial A_k}{\partial \mathbf{r}_i}
=
\sum_k W_k p_k \frac{\partial A_k}{\partial \mathbf{r}_i} \, ,
\]
which is the same expression as that for the gradient of $\Delta G_{\rm
cav}$ (see Appendix A of reference \cite{Gallicchio:Levy:2004}) with the replacement
\[
\gamma_k \rightarrow \frac{1}{4 \pi} W_k p_k
\]
The corresponding expression for the gradient of $\Delta G_{\rm GB}$
follows from the substitution:
%When including the surface area filter function:
%\[
%A_k = f_A(A_k) A_k
%\]
%we have:
%\begin{equation}
%\left( \frac{\partial \Delta G_{\rm vdW}}{\partial \mathbf{r}_i}
%\right)_{Q3} = 
%\sum_k \frac{f_A(A_k) + f'_A(A_k) A_k}{4 \pi} 
%p_k W_k \frac{\partial \bar{A}_k}{\partial \mathbf{r}_i} \, . 
%\label{dGvdw_dri3} 
%\end{equation}
%Therefore to compute this component of the gradient of $\Delta G_{\rm
%  vdW}$ it is sufficient to follow the same steps used to compute the
%  gradient of $\Delta G_{\rm cav}$ outlined in the 2004 paper. Again,
%  the corresponding expression for the gradient of $\Delta G_{\rm GB}$
%  follows from the usual $W_k \rightarrow u_\epsilon U_k$ substitution:
\[
\gamma_k \rightarrow \frac{1}{4 \pi} U_k p_k
\]

\subsection{Derivatives of HB Correction Energy}

From Eq.~(\ref{HB_corr_energy}) we have
\begin{equation}
\frac{\partial \Delta G_{\rm HB}}{\partial \mathbf{r}_i} =
\sum_s h_s S'(w_s) \frac{\partial w_s}{\partial \mathbf{r}_i} \, .
\label{GHB_der_1}
\end{equation}
Inserting Eqs.~(\ref{s_occupancy}) and (\ref{s_free_volume}) in Eq.\
(\ref{GHB_der_1}) gives
\begin{equation}
\frac{\partial \Delta G_{\rm HB}}{\partial \mathbf{r}_i} =
- \sum_{sj} \frac{h_s S'(w_s)}{V_s} \frac{\partial V_{sj}}{\partial
  \mathbf{r}_i}
+ \sum_{s,j<k} \frac{h_s S'(w_s)}{V_s} 
\frac{\partial V_{sjk}}{\partial  \mathbf{r}_i}
- \ldots
\end{equation}
where
\begin{equation}
\frac{\partial V_{sjk\ldots}}{\partial \mathbf{r}_i} =
 \left( \frac{\partial V_{sjk\ldots}}{\partial \mathbf{r}_i} 
 \right)_{\mathbf{r}_s} + 
\frac{\partial \mathbf{r}_s}{\partial \mathbf{r}_i} 
\frac{\partial V_{sjk\ldots}}{\partial  \mathbf{r}_s}
\end{equation}
where the first term on the r.h.s.\ represents the derivative of the
overlap volume with respect to the position of atom $i$ keeping the
position of the water site $s$ fixed, and the second term reflects the
change of overlap volume due to a variation of the position of the
water site caused by a shift in position of atom $i$. The latter term
is non-zero only if $i$ is one of the parent atoms of the water site.

\newpage

\providecommand{\refin}[1]{\\ \textbf{Referenced in:} #1}

%\bibliography{solvation}
%\bibliographystyle{unsrt}

\end{document}